\documentclass[%
 reprint,
 nofootinbib,
 amsmath,amssymb,
 aps,
]{revtex4-1}

\usepackage{hyperref}% add hypertext capabilities
\hypersetup{
    colorlinks,
    linkcolor={red!50!black},
    citecolor={blue!50!black},
    urlcolor={blue!80!black},
}

\usepackage{aas_macros}
\usepackage{natbib}
\usepackage{color}
\usepackage{mathtools,amssymb} % includes amsmath
\usepackage{graphicx,psfrag,subfigure}
\usepackage{sidecap}
\usepackage{soul,float}
\usepackage{algorithm,algpseudocode}
\usepackage{xcolor}
\usepackage{enumitem}
\usepackage{gensymb}
\usepackage{xspace}
\usepackage{nicefrac}
\usepackage{relsize}
\usepackage[utf8x]{inputenc}
\usepackage{fontawesome}
\usepackage{fancyvrb}
\usepackage{tablefootnote}
\usepackage{empheq}
\usepackage{lmodern}

\DeclareUnicodeCharacter{981}{$\phi$}

\DeclareMathAlphabet{\mathbbgreek}{U}{bbold}{m}{n}

\newcommand{\MUSE}{MUSE\xspace}

\newcommand{\Aphi}{A_{\phi}}
\newcommand{\Aphib}[1][]{A_{\phi}^{\ifthenelse{\equal{#1}{}}{b}{#1}}}
\newcommand{\hAphi}{\hat A_{\phi}}
\newcommand{\hAphib}[1][]{\hat A_{\phi}^{\ifthenelse{\equal{#1}{}}{b}{#1}}}

\newcommand{\AEb}{A_{E}^b}
\newcommand{\hAEb}{\hat A_{E}^b}

\newcommand{\op}[1]{\mathbb{#1}}
\newcommand{\Len}[1][]{\op L\ifthenelse{\equal{#1}{}}{}{(#1)}}
\newcommand{\Cflen}[1][]{\op{\widetilde{C}}_{f}\ifthenelse{\equal{#1}{}}{}{(#1)}}
\newcommand{\Cf}[1][]{\op C_{f}\ifthenelse{\equal{#1}{}}{}{(#1)}}
\newcommand{\Cphi}[1][]{\op C_{\phi}\ifthenelse{\equal{#1}{}}{}{(#1)}}
\newcommand{\Cn}{\op C_{n}}
\newcommand{\opG}[1][]{\op G\ifthenelse{\equal{#1}{}}{}{(#1)}}
\newcommand{\D}[1][]{\op D\ifthenelse{\equal{#1}{}}{}{(#1)}}

\newcommand{\smuse}[1][]{s^{\rm \scriptscriptstyle \MUSE}\ifthenelse{\equal{#1}{}}{}{_#1}}
\newcommand{\gmuse}[1][]{g^{\rm \scriptscriptstyle \MUSE}\ifthenelse{\equal{#1}{}}{}{_#1}}
\newcommand{\hmuse}[1][]{h^{\rm \scriptscriptstyle \MUSE}\ifthenelse{\equal{#1}{}}{}{_{#1}}}
\newcommand{\smap}[1][]{s^{\rm \scriptscriptstyle MAP}\ifthenelse{\equal{#1}{}}{}{_#1}}
\newcommand{\hatthetamuse}{\hat \theta^{\rm \scriptscriptstyle \MUSE}}
\newcommand{\x}{x}
\newcommand{\zMAP}{\hat z_{\theta,\x}}

\newcommand{\configSMALL}{\textsc{CMBS4-150d}\xspace}
\newcommand{\configBIG}{\textsc{SPT-3G}\xspace}

\begin{document}

\title{MUSE: Marginal Unbiased Score Expansion and Application to CMB Lensing}

\author{Marius Millea}
\email{mariusmillea@gmail.com}
\affiliation{Department of Physics, University of California, Berkeley, CA 94720, USA}
\affiliation{Department of Physics, University of California, Davis, CA 95616, USA}

\author{Uro\v s Seljak}
\affiliation{Department of Physics, University of California, Berkeley, and Lawrence Berkeley National Laboratory, Berkeley, CA 94720, USA}

\begin{abstract}
We present the marginal unbiased score expansion (\MUSE) method, an algorithm for generic high-dimensional hierarchical Bayesian inference. \MUSE performs approximate marginalization over arbitrary non-Gaussian latent parameter spaces, yielding Gaussianized asymptotically unbiased and near-optimal constraints on global parameters of interest. It is computationally much cheaper than exact alternatives like Hamiltonian Monte Carlo (HMC), excelling on funnel problems which challenge HMC, and does not require any problem-specific user supervision like other approximate methods such as Variational Inference or many Simulation-Based Inference methods. \MUSE makes possible the first joint Bayesian estimation of the delensed Cosmic Microwave Background (CMB) power spectrum and gravitational lensing potential power spectrum, demonstrated here on a simulated data set as large as the upcoming South Pole Telescope 3G 1500\,deg$^2$ survey, corresponding to a latent dimensionality of ${\sim}\,6$ million and of order 100 global bandpower parameters. On a subset of the problem where an exact but more expensive HMC solution is feasible, we verify that \MUSE yields nearly optimal results. We also demonstrate that existing spectrum-based forecasting tools which ignore pixel-masking underestimate predicted error bars by only ${\sim}\,10\%$. This method is a promising path forward for fast lensing and delensing analyses which will be necessary for future CMB experiments such as SPT-3G, Simons Observatory, or CMB-S4, and can complement or supersede existing HMC approaches. The success of \MUSE on this challenging problem strengthens its case as a generic procedure for a broad class of high-dimensional inference problems.
\end{abstract}

\maketitle
% \tableofcontents

\section{Introduction}

Bayesian inference is a highly successful paradigm for quantifying uncertainty in the face of observed data. The procedure centers on updating a prior probability distribution, $\mathcal{P}(\theta)$, with the likelihood of some observed data, $\mathcal{P}(\x\,|\,\theta)$, where $\theta$ represent some parameters of interest and $x$ represents the data. Bayes' Theorem describes the form of this update:
\begin{align}
    \label{eq:bayes}
    \mathcal{P}(\theta\,|\,\x) = \frac{\mathcal{P}(\x\,|\,\theta) \, \mathcal{P}(\theta)}{\mathcal{P}(\x)}.
\end{align}
The posterior, $\mathcal{P}(\theta\,|\,\x)$, then summarizes the entirety of the information on $\theta$ provided by the observations. 
%In the hybrid Bayesian-frequentist approach, the likelihood can be used to constract an estimator for $\theta$, the marginal maximum likelihood estimate (MMLE\footnotemark), which is ``optimal'' in the sense that it is asymptotically unbiased and minimum-variance given appropriate regularity conditions.

%\footnotetext{Often this is called just the maximum  (MLE), but we will be refer to is as the MMLE since it maximizes the likelihood marginalized over the latent space variables, as opposed maximizing the joint likelihood of latent space variables and of parameters of interest, introduced in Eqn.~\eqref{eq:likelihood}.}

In many problems of interest, the data do not depend just on $\theta$, but also on a set of unobserved latent variables, $z$, which themselves depend stochastically on $\theta$. In this case, the likelihood in Eqn.~\eqref{eq:bayes} involves a marginalization over the latent space $z$,
\begin{align}
    \label{eq:likelihood}
    \mathcal{P}(\x\,|\,\theta) = \int \!{\rm d}^N\!z \; \mathcal{P}(\x,z\,|\,\theta) = \int \!{\rm d}^N\!z \; \mathcal{P}(\x\,|\,z,\theta)\mathcal{P}(z\,|\,\theta),
\end{align}
where $\mathcal{P}(\x,z\,|\,\theta)$ is the joint likelihood of both data and of $z$, given $\theta$. These problems are considered ``hierarchical'' Bayesian problems, as there can be a hierarchy of latent variables, each depending probabilistically on the previous set. For the purposes of our work, we will consider $z$ to encompass the entire latent space, defined as all parameters other than those of interest, $\theta$. 

The fundamental challenge of hierarchical Bayesian inference is performing the integral in Eqn.~\eqref{eq:likelihood}. Closed-form solutions rarely exist for all but the simplest problems, and the latent space can often be very high-dimensional and non-Gaussian, making numerical integration costly or intractable. 

Some intermediate quantities which are easy to calculate can be useful for diagnosis or as pieces of other algorithms, but by themselves are not helpful for inferring parameters. For example, maximizing the joint likelihood or joint posterior of $(\theta,z)$ to produce the joint maximum likelihood estimate (JMLE) or joint maximum a posteriori estimate (JMAP), respectively, do not give useful estimates of $\theta$ since they only represent the peak of the integrand in Eqn.~\eqref{eq:likelihood} but have not performed the necessary integral. Attempting to use the JMAP or JMLE yields estimates of $\theta$ which are biased or not meaningful \citep[e.g.][]{neyman1948,millea2019}. Only the marginal maximum likelihood estimate (MMLE) of $\theta$, which maximizes $\mathcal{P}(x\,|\,\theta)$, is asymptotically unbiased, but of course still involves the difficult integral in Eqn.~\eqref{eq:likelihood}.

Several popular solutions to the marginalization problem exist, each with various advantages and tradeoffs. Methods such as Hamiltonian Monte Carlo (HMC) are asymptotically exact, but can become very slow for high-dimensional problems or even for moderate dimensions with sufficiently non-Gaussian latent spaces (see e.g.\,\,\cite{betancourt2017} for a review). Variational Inference (VI) forms another popular class of methods, which are deemed to be faster than HMC. However, VI is approximate and generically requires the user to chose a family of variational distributions, rendering the method less automatic (see e.g.\,\,\cite{zhang2018} for a review). For cosmological applications, simplifications such as mean-field VI are inaccurate due to the correlations between the modes induced by survey masks, while full rank VI is equally infeasible due to the high dimensionality of the problem, which would require estimation of a very high dimensional covariance matrix. The difficulty of evaluating the marginal likelihood of Eqn.~\eqref{eq:likelihood} has popularized the use of methods where the likelihood is not used at all, which go under the name of Likelihood Free Inference (LFI) or Simulation-Based Inference (SBI). These have attracted recent interest, but struggle for data and/or latent spaces which are very high-dimensional (see e.g.\,\,\cite{cranmer2020} for a review). 

Here, we present the Marginal Unbiased Score Expansion (\MUSE) method. It provides an often highly accurate approximation to Eqn.~\eqref{eq:likelihood} (or rather, to its gradient, as we will discuss), and is much faster to compute than exact methods. The approximation was developed by \citet{seljak2017}, with some applications to Gaussian problems in \citet{horowitz2019}. In this work, we extend previous results by quantifying the properties of this approximation for general non-Gaussian latent spaces, rendering it applicable to a much wider range of problems\footnote{We also dub this generalized version ``MUSE''.}. 

We show that regardless the structure of the latent space, \MUSE provides estimates of parameters which are asymptotically unbiased, meaning they are unbiased as long as many different data modes contribute to constraints on each $\theta$. This condition is quite often met automatically for the high-dimensional problems where \MUSE is useful over alternatives anyway. In the context of cosmology, this is typical, as one often seeks to infer constraints on a small number of parameters or bandpowers from the aggregate of a huge number of data modes. 

Additionally, as both a demonstration and as a novel solution in its own right, we apply \MUSE to the challenging problem of Cosmic Microwave Background (CMB) gravitational lensing (see e.g.\,\,\cite{lewis2006} for a review). Specifically, we use it to infer parameters and/or bandpowers of the gravitational lensing potential joint with the unlensed CMB. This problem is hierarchical because the parameters of interest control the statistics of the unobserved (latent) unlensed CMB and lensing potential. It is non-Gaussian due to the nature of lensing distortions. And it is high-dimensional because the size of the latent space is on the order of the number of map pixels, which can be millions for modern observations. As such, it is a perfect test-bed for \MUSE.

Traditionally, analysis of CMB lensing has relied on the so-called ``quadratic estimator,'' (QE), which is an estimator formed from quadratic combinations of the data \citep{zaldarriaga1999,hu2002}. The QE is near-optimal for present day instrumental noise levels, and is based on an explicit semi-analytic formula which does not involve marginalization over any latent space. It has been widely used in almost all CMB lensing analyses to-date. However, as first shown by \citet{hirata2003,hirata2003a} and \citet{seljak2004}, at noise thresholds which are currently being crossed by the most sensitive experiments, the QE ceases to be near-optimal and Bayesian methods which fully extract all-orders information from the data can yield significantly better results. At the noise levels of the planned CMB-S4 experiment \cite{abazajian2016}, this includes reconstructing the gravitational lensing field to ${\sim}\,10$ times lower noise levels \cite{millea2020a} and yielding delensed maps of $B$ modes which allow ${\sim}\,3$ times better constraints on the amplitude of primordial gravitational waves, $r$ \citep{carron2018}.

The original work by \citet{hirata2003,hirata2003a} gave a method applicable to idealized mask-free data and set the stage for a number of ensuing improvements. The issue of masking, a necessity for any real analysis which must excise contaminated or unobserved parts of the sky, is central to the challenge of optimal lensing. While it is easy by comparison to devise estimates which assume full-sky or periodic flat-sky boundary conditions without masking, the impact of masking is drastic as it transforms the correlation structure of the high-dimensional latent space from sparse to dense, causing the breakdown of many simple approximations which could otherwise be used. \citet{carron2017} extended the original work in \cite{hirata2003,hirata2003a} and computed a MAP estimate of $\phi$ for realistic data conditions which included masking, but did not attempt the integral in Eqn.~\eqref{eq:likelihood}, which would be needed to infer parameter constraints. A power spectrum estimate based on this MAP was recently given in \citet{legrand2021}, but the approximations are not validated in the presence of masking. \citet{carron2018} also used the estimate to perform a brute-force integration of Eqn.~\eqref{eq:likelihood} in the case that $\theta$ is one-dimensional, but the method does not scale to higher dimensions. Machine-learning estimates of $\phi$ have been demonstrated by \citet{caldeira2018} and \citet{guzman2021}, but which likewise do not attempt the integral in Eqn.~\eqref{eq:likelihood}. All of these extensions also at present lack a demonstrated way to infer power spectra of the unlensed CMB, needed for full parameter extraction. Full-sky tools have been developed by \citet{green2016} and \citet{hotinli2021} which consider optimal joint lensing reconstruction and delensing, and while they serve as very useful power spectrum forecasting tools, they do not correspond to a map-level procedure which could be applied to real data 

To-date, the only tractable method for performing the latent-space integration while considering lensing, delensing, and realistic data conditions, is based on the HMC sampling procedure of \citet{anderes2015,millea2019,millea2020b}. This method was applied to South Pole Telescope data to demonstrate the first joint parameter estimation from an optimally reconstructed $\phi$ field and delensed CMB by \citet{millea2020a}. While the HMC lensing approach has several appealing features, its downside is that it is slow and the Monte Carlo sampler is necessarily sequential and cannot be trivially sped up. For example, the 100\,deg$^2$ of polarization data analyzed by \cite{millea2020a} took roughly 4 wall-hours on GPU, with a naive scaling to the entire SPT-3G 1500\,deg$^2$ survey suggesting HMC chains would require a week to converge. Here we will demonstrate an analysis of this 1500\,deg$^2$ dataset which completes in hours, and which is very amenable to trivial parallelization. The output of the estimate is a set of $\phi$ and delensed $E$-mode bandpowers as well as their joint covariance, making this a familiar data product for cosmologists to then use in a subsequent parameter estimation step. 

Alongside this paper, we provide a software package, \textsc{MuseInference.jl}\footnote{\href{https://github.com/marius311/MuseInference.jl}{https://github.com/marius311/MuseInference.jl}}, which is a generic implementation of \MUSE that can be used on any hierarchical Bayesian problem. This package has an interface into the probabilistic programming language (PPL) \textsc{Turing.jl} \cite{ge2018}, and can immediately be applied to existing models and compared against HMC or VI. Interfaces to other PPLs like \textsc{Stan} or \textsc{PyMC} are planned. The interface also includes the existing \textsc{CMBLensing.jl}\footnote{{\href{https://github.com/marius311/CMBLensing.jl}{https://github.com/marius311/CMBLensing.jl}}} code for application to the CMB lensing problem.

We begin in Sec.~\ref{sec:mpm} with a description of the \MUSE procedure aimed at a general audience, and demonstrate it on a generic toy problem in Sec.~\ref{sec:toympm}. In Sec.~\ref{sec:lensingposterior} we introduce the CMB lensing problem and present our main lensing results, before summarizing and giving concluding comments in Sec.~\ref{sec:conclusion}.

\section{The algorithm}
\label{sec:mpm}

We begin with a generic description of the \MUSE algorithm, applicable to any hierarchical Bayesian problem. In the later sections, we will turn to our specific application of CMB gravitational lensing.

\subsection{The \MUSE approximation}

The \MUSE algorithm is based on an approximation to the gradient of the marginal log-likelihood, a quantity usually called the marginal score:
\begin{equation}
    s_i(\theta,\x) \equiv \frac{d}{d\theta_i} \log \mathcal{P}(\x\,|\,\theta)
\end{equation}
The marginal score represents a lossless compression of the data, containing all information on parameters which can theoretically be extracted \citep{alsing2018}. While many approximate Bayesian methods seek to approximate the value of the marginal or joint posterior, approximating the marginal score is just as good, and potentially more direct if in the end we are only interested in inferences of $\theta$ anyway. For example, if we had access to the marginal score, we could use exact Bayesian methods such as HMC to infer $\theta$ directly, since the relevant Hamiltonian trajectories would depend on just the marginal score.\footnotetext{Although typical HMC implementations would also need the value of the posterior for the error-correcting Metropolis-Hastings step, this is only a practical issue of symplectic integration error, and is not theoretically needed in the limit of infinitely small step-size.} Similarly, optimal estimators like the MMLE can be defined entirely in terms of the marginal score, since the estimate is the parameter vector, $\hat \theta^{\rm MMLE}$, which solves
\begin{equation}
    \label{eq:mmle}
    s_i(\hat \theta^{\rm MMLE},\x) = 0
\end{equation}
It is thus well-motivated to search for efficient ways to compute or approximate the marginal score.

Of course, the exact marginal score still requires performing the difficult integral in Eqn.~\eqref{eq:likelihood}. The MUSE solution involves an approximation which is extremely fast to compute in comparison to exact integration. As motivation, consider a Taylor series expansion of the joint likelihood,
\begin{align}
    \label{eq:museapprox}
    \log\, & \mathcal{P}(\x,z\,|\,\theta) = \\ 
    &= \log \mathcal{P}(\x,\zMAP\,|\,\theta) + \frac{1}{2} (z-\zMAP)^\dagger H_{\theta,\x} (z-\zMAP) + ..., \nonumber
\end{align}
where $\zMAP$ is the maximum a posteriori (MAP) estimate of the latent space variables given $\x$ and fixed $\theta$, 
\begin{align}
    \label{eq:zmap}
    \zMAP \equiv \underset{z}{\rm argmax} \; \log \mathcal{P}(\x,z\,|\,\theta),
\end{align}
and $H_{\theta,\x}$ is the Hessian matrix at this point.
In terms of this expansion, the marginal score is then,
\begin{multline}
    s_i(\theta,\x) = s_i^{\rm MAP}(\theta,\x) \; + \\ \frac{d}{d\theta_i} \log \int {\rm d}^N\!z \, \exp \left[ \frac{1}{2} \, (z-\zMAP)^\dagger H_{\theta,x} (z-\zMAP) + ... \right]
    \label{eq:smuse_partial}
\end{multline}
where we have defined the gradient evaluated at the MAP as, 
\begin{align}
    s_i^{\rm MAP}(\theta,\x) \equiv \frac{d}{d\theta_i}\log \mathcal{P}(\x,\zMAP\,|\,\theta).
\end{align}
Note that the chain rule term which would appear above involving $d\zMAP/d\theta$ never needs to be computed because it is multiplied by the gradient of the distribution, which is by definition zero at the MAP.

One common approach for approximating Eqn.~\eqref{eq:smuse_partial} is to keep only the quadratic term in the exponential, yielding a Gaussian integral with an analytic solution. This is the well-known Laplace approximation. However, this still requires obtaining the trace and inverse of the Hessian matrix, which in practice may also be extremely difficult due to the high dimensionality of the latent space and hence of this matrix. The key insight of \MUSE is not to attempt to perform the remaining integral at all, but rather approximate it with its data-averaged value. This can in turn easily be obtained from the ``unbiased score equation,'' which is the fact that any arbitrary score function (under regularity conditions) obeys
\begin{equation}
    \Big \langle s_i(\theta,\x) \Big \rangle_{\x \sim \mathcal{P}(\x\,|\,\theta)} = 0.
\end{equation}
Enforcing this condition and solving for the expected value of the integral yields the \MUSE marginal score approximation \cite{seljak2017}:
\begin{align}
    \label{eq:smuse}
    \boxed{\smuse[i](\theta,\x) \equiv \smap[i](\theta,\x) - \Big \langle \smap[i](\theta,\x^\prime) \Big \rangle_{\x^\prime \sim \mathcal{P}(\x^\prime\,|\,\theta)}.}
\end{align}
In practice, the second term in Eqn.~\eqref{eq:smuse} is computed via a Monte Carlo average over a suite of forward data simulations generated at the given value of $\theta$.

Note that the \MUSE approximation has the desirable property that in the limiting case of a Gaussian joint likelihood, where the latent space is Gaussian with a data-independent Hessian, it becomes exact (an explicit example of this is given in Appendix \ref{app:exactmpm}). Even for mildly non-Gaussian latent spaces, one expects the data-dependence of the integral to be small, with most of the data-dependence instead captured by the MAP term. Additionally, regardless of whether the latent space is Gaussian or not, $\smuse$ always obeys the unbiased score equation by construction. This feature will turn out to be key allowing Bayesian or frequentist estimates built from $\smuse$ to remain unbiased.

We emphasize that the \MUSE approximation does not correspond to the Laplace approximation for the joint likelihood. In fact, it may not correspond to {\it any} approximation for the joint likelihood, because $\smuse$ is not, in general, a conservative vector field. This means it cannot be written as the gradient of some scalar function, which would then be interpretable as the approximate distribution. Exceptions to this include the Gaussian problem, where the non-conservative terms in $\smuse$ cancel, and the case of a one dimensional $\theta$, where this distinction does not exist. This technical detail will have a few important consequences, discussed below.

\subsection{The frequentist view}
\label{sec:freqmuse}

We now describe both a frequentist and Bayesian approach for parameter inference which make use of $\smuse$. Ultimately, both correspond to performing the identical computation, and differ only in interpretation. This is not surprising since the regime where \MUSE is best applicable is where the $\theta$ are well-constrained relative to the prior, and hence where Bayesian and likelihood-based frequentist methods agree. It is instructive, however, to follow the assumptions inherent in each description, which might point to different types of future extensions of the method. We begin with the frequentist version.

In the frequentist approach, we are interested in building an estimator for $\theta$. In analogy to the MMLE defined in Eqn.~\eqref{eq:mmle}, it is natural to define the \MUSE estimate as the parameter vector which solves, 
\begin{equation}
    \label{eq:museestimate}
    \boxed{ \smuse[i](\hatthetamuse,\x) = 0. }
\end{equation}
In this way, if the latent space is Gaussian where \MUSE is exact, we recover the MMLE, and \MUSE is therefore asymptotically unbiased and minimum variance.

Note that while Eqn.~\eqref{eq:mmle} can be rephrased as maximizing a scalar function (i.e., maximizing the marginal likelihood), the fact that $\smuse$ may be non-conservative means \MUSE must generically be regarded as a vector-valued root-finding problem. Such problems are not guaranteed to have a solution, and if no solution is found for some particular case, then \MUSE cannot be used. Having noted this, we have not found it to be a typical concern except in some pathological instances.

We next need to determine the bias and covariance of the \MUSE estimator. We will consider the asymptotic limit of a large number of observations, $N$, since we are targeting problems where a large $N$ has driven the estimator distribution towards Gaussian by the central limit theorem. The log-likelihood of $N$ data, $x_1, ..., \x_N$, drawn independently from $\mathcal{P}(\x\,|\,\theta)$, is the sum of the log-likelihoods of each. Accordingly, the \MUSE estimate for $N$ data is implicitly defined by the solution to 
\begin{multline}
    \smuse[i](\hat \theta,\{\x_n\}) \equiv \\ \frac{1}{N}\sum_{n=1}^{N} \smap[i](\hat \theta,\x_n) - \Big \langle \smap[i](\hat \theta,x^\prime) \Big \rangle_{x^\prime \sim \mathcal{P}(x^\prime\,|\,\hat\theta)}=0.\label{eq:mpm_implicit_asymptotic}
\end{multline}
where we have defined the \MUSE gradient for $N$ data as $\smuse[i](\hat \theta,\{\x_n\})$ for later use, and have used $\hat\theta\,{\equiv}\,\hatthetamuse$ for brevity. In the limit of $N\,{\rightarrow}\,\infty$, the summation in Eqn.~\eqref{eq:mpm_implicit_asymptotic} becomes an expectation value over $\x \,{\sim}\,\mathcal{P}(\x\,|\,\theta)$, and the equation is trivially solved when $\hat \theta\,{=}\,\theta$, demonstrating that the \MUSE estimate is asymptotically unbiased (for any Gaussian or non-Gaussian latent space). 

Note that for finite $N$, the difference between the sum in the term in brackets in Eqn.~\eqref{eq:mpm_implicit_asymptotic} and its asymptotic limit will scale as $1/\sqrt{N}$ by the central limit theorem, thus the estimator bias scales like $1/\sqrt{N}$ times its standard deviation, similarly as for the MMLE. We also note that the \MUSE estimate is trivially unbiased for {\it any} $N$ when the $\theta$ which generates the simulations in Eqn.~\eqref{eq:smuse} happens to be the truth. Thus, if some prior knowledge suggests some regularization which brings the \MUSE estimate closer to truth\footnote{In the context of cosmology, one such regularization arises if the $\theta$ represent a spectral density which is expected from physical arguments to vary slowly with scale, such as the CMB or matter power spectrum. In this case, $\theta$ can be regularized by applying some chosen smoothing kernel.}, the bias can actually be smaller than for the MMLE.

Finally, we consider the covariance of the \MUSE estimate. We first note that it is straightforward to compute the estimator covariance at some $\theta$ via Monte Carlo, by running the \MUSE estimate on a suite of simulated data and taking the empirical covariance. If the computational cost of running a sufficient number of simulations is not prohibitive, this is likely the easiest approach in practice, and is guaranteed to give an exact answer up to Monte Carlo errors. We can, however, significantly reduce the computational cost. If the latent space is known to be Gaussian, \citet{horowitz2019} demonstrated a fast and exact approach. For the non-Gaussian latent spaces of interest in this work, we supersede the suggestions in \citet{seljak2017} with one which works for more general non-Gaussian distributions.

First, expand Eqn.~\eqref{eq:mpm_implicit_asymptotic} to first order around the true value, denoted $\theta^\ast$,
\begin{equation}
    \smuse[i](\theta^\ast,\{\x_n\}) \, + \, (\hat \theta_j - \theta^\ast_j) \, \hmuse[ij] (\theta^\ast,\{\x_n\}) = 0
\end{equation}
where we have defined the Jacobian
\begin{align}
    \hmuse[ij] = \frac{d\smuse[i]}{d\theta_j}.
\end{align}
Introducing a factor of $N$ and rearranging terms yields
\begin{multline}
    \sqrt{N}(\hat \theta_j - \theta^\ast_j) = \\ - \left[ \frac{1}{N} \hmuse[ij] (\theta^\ast,\{\x_n\}) \right]^{-1} \left[ \sqrt{N}\frac{1}{N} \smuse[i](\theta^\ast,\{\x_n\}) \right].
\end{multline}
Assuming suitable regularity conditions, the second term in brackets can be shown by the central limit theorem to converge in probability as $N\rightarrow\infty$ to a normal distribution with zero mean and with covariance given by $J_{ij}$, and the quantitiy in the first term in brackets to converge to $H_{ij}$, where these matrices are defined as 
\begin{align}
    J_{ij} &= \Big\langle \smuse[i](\theta^\ast,\{\x_n\}) \, \smuse[j](\theta^\ast,\{\x_n\}) \Big\rangle_{\x_n\overset{\rm iid}{\sim}\,\mathcal{P}(\x\,|\,\theta^\ast)} \label{eq:defJ} \\
    H_{ij} &= \Big \langle \hmuse[ij] (\theta^\ast,\{\x_n\}) \Big \rangle_{\x_n\overset{\rm iid}{\sim}\,\mathcal{P}(\x\,|\,\theta^\ast)}. \label{eq:defH}
\end{align}
These expressions can be further simplified to be written only in terms of averages of gradients at the MAP (see Appendix~\ref{app:JH}):
\begin{empheq}[box=\fbox]{align}
    J_{ij} &= \Big\langle \smap[i](\theta^\ast,\x) \, \smap[j](\theta^\ast,\x) \Big\rangle_{\x\sim\mathcal{P}(\x\,|\,\theta^\ast)} \\ & \quad -\Big\langle \smap[i](\theta^\ast,\x) \Big\rangle \Big \langle \smap[j](\theta^\ast,\x) \Big\rangle_{\x\sim\mathcal{P}(\x\,|\,\theta^\ast)} \nonumber \\
    H_{ij} &= \left. \frac{d}{d\theta_j} \left[ \Big\langle \smap[i](\theta^\ast,\x) \Big\rangle_{\x\sim\mathcal{P}(\x\,|\,\theta)} \right] \right|_{\theta=\theta^\ast}
    \label{eq:JH}
\end{empheq}
The interpretation of $J$ is straightforward: it is the covariance of the MAP gradient simulations. As these are already computed for the purposes of computing $\smuse$ itself, they do not add any extra computational cost. $H$ involves one extra derivative of these gradient simulations, and we note that careful attention should be given to which variables are held fixed; the derivative does not act on the argument of $\smap$, which is held fixed at $\theta^\ast$, instead only acting on the parameters which generate the simulated data, $\x$. Supposing that finite differences were used to compute $H$ highlights another interpretation of this term: $H$ is computed by injecting infinitesimal parameter shifts into simulated data, and observing how the MAP gradient changes in response.

As long as $H$ is invertible, the continuous mapping theorem then gives the final result for the asymptotic covariance of the \MUSE estimate:
\begin{align}
    \Sigma_{ij}^{\rm \MUSE} \equiv \langle \Delta \hatthetamuse_i \Delta  \hatthetamuse_j \rangle = (H^{-1} J \, H^{-\dagger})_{ij}
    \label{eq:asympcov}
\end{align}
Note that one can show that for Gaussian problems, $J\,{=}\,H\,{=}\,\mathcal{F}^{-1}$ where $\mathcal{F}$ is the Fisher information matrix, such that the estimator saturates the Cram\'er-Rao bound and is considered optimal. This follows from the fact that for Gaussian problems, \MUSE becomes the MMLE, which itself is known to asymptotically saturate this bound.

\subsection{The Bayesian view}

In the Bayesian approach, instead of a point estimate, we are interested in exploring the posterior distribution, $\mathcal{P}(\theta\,|\,\x)$. Although we can use the \MUSE score in conjunction with the prior to approximate the gradient of the log-posterior, 
\begin{align}
    \gmuse[i](\theta,x) \equiv \smuse[i](\theta, x) + \frac{d}{d\theta_i}\log\mathcal{P}(\theta),
\end{align}
the fact that $\smuse$ and hence $\gmuse$ are not guaranteed to be conservative vector fields means there is no trivial way to use this to back out an approximation to $\mathcal{P}(\theta\,|\,\x)$. Accordingly, HMC is not in theory applicable to $\gmuse$ because the Hamiltonian equations underpinning HMC assume two trajectories starting and ending at the same points in parameter space yield the same change in conjugate momenta, which would not be the case for a non-conservative $\gmuse$. It may be the case that this is not a problem in practice, or that the Hamiltonian dynamics can be modified to account for it, but we have not explored this avenue. 

One option to proceed is a Gibbs sampling approach, considering one parameter at a time. In one dimension, issues of conservativeness are irrelevant, and we can integrate $\smuse$ on a grid to obtain a full posterior shape, draw a sample, then continuously repeat for subsequent dimensions. We give an example of the grid evaluation in the next section, but have not pursued this further.

Another natural Bayesian approach is to consider $\hatthetamuse$ as a summary statistic, and explore the posterior $\mathcal{P}(\theta\,|\,\hatthetamuse)$. In summarizing the data with $\hatthetamuse$, we have potentially lost some information, and this will be reflected in a potentially less constraining $\mathcal{P}(\theta\,|\,\hatthetamuse)$ as compared to $\mathcal{P}(\theta\,|\,\x)$. However, this will only be significant given large latent non-Gaussianity, and regardless, the posterior inference will be valid.

If we suspect that $\mathcal{P}(\theta\,|\,\hatthetamuse)$ is considerably non-Gaussian, we could use any of a number of SBI methods to map out this distribution \cite[e.g.][]{cranmer2020}. In this sense, we can view $\hatthetamuse$ as a near-optimal data-compression step, of the kind required by many SBI methods. Because \MUSE is fast to compute, the total computational cost can still be well below the cost of performing full HMC or SBI on the joint posterior. 

If we instead assume a near-Gaussian $\mathcal{P}(\theta\,|\,\hatthetamuse)$, we can forego SBI and compute the distribution under some simple assumptions. Writing $\hat\theta\equiv \hatthetamuse$ for clarity, the posterior conditioned on the summary statistic is
\begin{align}
    \mathcal{P}(\theta\,&|\,\hat \theta) = \frac{\mathcal{P}(\hat \theta\,|\,\theta) \, \mathcal{P}(\theta)}{\mathcal{P}(\hat \theta)}.
\end{align}
The first term in the numerator can be written as
\begin{align}
    \mathcal{P}(\hat \theta\,|\,\theta) &= \int {\rm d}^N\!\x \; \mathcal{P}(\hatthetamuse \, | x) \mathcal{P}(x\,|\,\theta) \\
    &= \int {\rm d}^N\!\x \; \delta^P\big(\smuse(\hat\theta,x)\big) \mathcal{P}(x\,|\,\theta)
\end{align}
where $\delta$ is the Dirac delta function, $P$ is the dimensionality of $\theta$ and $N$ the dimensionality of $x$. We can use the first and second moments to approximate the mean and covariance of this Gaussian. The mean is,
\begin{align}
    \mu_i = \int {\rm d}^P \hat\theta \; \hat\theta_i \, \int {\rm d}^N\!\x \; \delta^P\big(\smuse(\hat\theta,x)\big) \mathcal{P}(x\,|\,\theta).
\end{align}
Assuming the likelihood is adequately peaked, it suffices to Taylor expand the \MUSE score around $\theta$, 
\begin{align}
    \smuse[i](\hat\theta,x) \approx \smuse[i](\theta,x) + (\hat\theta_j-\theta_j) \, \hmuse[ij](\theta,x)
\end{align}
then perform the integral over $\hat \theta$ to obtain
\begin{align}
    \label{eq:summarymean}
    \int {\rm d}^N\!\x \; \mathcal{P}(\x\,|\,\theta) \Big[ \hmuse[ji](\theta,\x)^{-1} \smuse[i](\theta,\x) + \theta \Big].
\end{align}
Assuming $\hmuse$ is independent of data and using the fact that $\smuse$ obeys the unbiased score equation yields simply $\mu=\theta$. We note that if desired, this assumption can be explicitly checked in practice since Eqn.~\eqref{eq:summarymean} is just an average over forward data simulations and can be computed via Monte Carlo.

The covariance will be
\begin{multline}
    \tilde\Sigma_{\ij}^{\rm MUSE} = \int {\rm d}^P \hat\theta \; (\hat\theta_i - \theta_i)(\hat\theta_j - \theta_j)  \\ \int {\rm d}^N\!\x \; \delta^P\big(\smuse[k](\hat\theta,\x)\big) \mathcal{P}(\x\,|\,\theta),
\end{multline}
where the tilde differentiates this covariance from the covariance of the \MUSE estimator defined in the previous section. Simplifying similarly as above yields
\\\begin{multline}
    \label{eq:summarycov}
    \int {\rm d}^N\!\x \; \mathcal{P}(\x\,|\,\theta) \Big[ \hmuse[ij](\theta,\x)^{-1} \smuse[j](\theta,\x) \\ \times \smuse[k](\theta,\x) \hmuse[kj](\theta,\x)^{-1}\Big],
\end{multline}
which can also be explicitly computed via Monte Carlo. Note that if we assume $\hmuse$ is realization-independent, then $\tilde\Sigma_{\ij}^{\rm MUSE}\,{=}\,\Sigma_{\ij}^{\rm MUSE}$, meaning the Bayesian and frequentist estimates are identical. 

With the Gaussian approximation to $\mathcal{P}(\hat \theta \, | \, \theta)$ ascertained in this way, one can combine it with any prior desired to obtain the full posterior $\mathcal{P}(\theta \, | \, \hat\theta)$. In the special case that the prior is also a Gaussian, $\mathcal{N}(\theta_{\rm p}, \Sigma_{\rm p})$, we have that
\begin{align}
    \mathcal{P}(\theta\,|\,\hat\theta) = \mathcal{N}\Big( \Sigma_{\rm tot}(\Sigma^{-1} \hat\theta + \Sigma_{\rm p}^{-1} \theta_{\rm p}),  \; \Sigma_{\rm tot}\Big).
    \label{eq:bayesmuse}
\end{align}
where $\Sigma_{\rm tot}\,{=}\,(\Sigma^{-1}\,{+}\,\Sigma^{-1}_{\rm p})^{-1}$. Note also that the mean of this distribution can be equivalently calculated by simply solving $\gmuse(\hat\theta,\x)\,{=}\,0$ rather than $\smuse(\hat\theta,\x)\,{=}\,0$. 

\subsection{Practical considerations}

In practice, either Bayesian or frequentist views of \MUSE involve first iteratively solving the vector equation $\gmuse[i](\hat\theta,\x)\,{=}\,0$ or $\smuse[i](\hat\theta,\x)\,{=}\,0$ for $\hat\theta$, respectively. We use Broyden's method \cite{broyden1965}, a standard choice for this type of problem. This requires an initial guess for the Jacobian, whose exact value would be $\hmuse$. We find it sufficient to approximate this as $J^{-1}$ or even ${\rm Diagonal}(J^{-1})$, which can be computed for free from simulations already performed for the first step. This choice impacts only the speed of convergence, not the final solution.

At each step of Broyden's method, we compute $\smuse$ at the current value of $\theta$. This involves generating $M$ data simulations given $\theta$, and finding the MAP solution of each. We note that this step is completely amenable to trivial parallelization, and on GPUs, is ideally performed efficiently with batching. These MAP solutions dominate the runtime of the algorithm, and the fact that they are parallelizable in this way is a particular strength of \MUSE. The MAP solutions can be found with standard methods like L-BFGS, although individual problems may have domain-specific solutions as well (e.g. the CMB lensing problem discussed later features an optimized solver based on coordinate descent). For performance, it is very advantageous to use the same random seeds for these simulations throughout each iterative Broyden step, and to start the MAP solver for each step from the solution for the previous step. Since $\theta$ is not changing much by the final Broyden iterations, the MAP solver will require very few steps. 

The choice of how many simulations, $M$, to use, is informed by how much Monte Carlo error one is willing to incur. The error between $\hat\theta$ computed using $M$ simulations and its true value in the limit $M\,{\rightarrow}\,\infty$, relative to its standard deviation, scales as $1/\sqrt{M}$. Thus, using 100 simulations incurs a possible error of $0.1\,\sigma$, which for many purposes is sufficient. Even 10 simulations yields an error of 0.3$\,\sigma$, which is often acceptable. If one is using \MUSE to compare the impact of two choices of modeling assumptions, one can use the same simulation random seeds for both runs, in which case the error cancels out to first order and the impact can be determined to much better than $0.1\,\sigma$ even with 100 simulations (and potentially even much fewer). Alternatively, if one is interested in the error between $\hat\theta$ and the true value of $\theta$, the scaling is instead $\sqrt{1+1/M}$. Thus, with only 10 simulations, the error relative to the true value is only increased by about $0.05\,\sigma$. Different choices of which metric to use to chose $M$ can be valid in different circumstances.

Computing the $J$ and $H$ matrices can vary in difficulty depending on the problem and depending on if second order Automatic Differentiation (AD) is available. The $J$ matrix can be computed from the same MAP solutions which went into the computation of $\smuse$ on the final Broyden iteration. If needed, additional simulations can be performed just at the final $\theta$ for the purposes of recovering $J$ and its inverse to better accuracy. Additionally, since $J$ is a covariance, shrinkage estimators can be used if some particular structure is expected. Computing $H$ involves propagating a second-order derivative through the MAP solver, which, with second-order AD, can be done in the same pass as the MAP solutions needed for $J$. If the AD library does not provide higher-order derivatives, finite differences can be used, which may still be quite fast as the dimensionality of $\theta$ is generally small. We also note that $H$ is typically quite realization-independent, thus does not require averaging over a large number of simulations (and this can of course be checked in practice). Finally, for many problems, particularly in cosmology where a viable fiducial model is already known, $J$ and $H$ can be computed once at the fiducial model and only need to be recomputed under significant changes to modeling assumptions.

\vfill

\begin{figure*}
    \includegraphics[width=\textwidth]{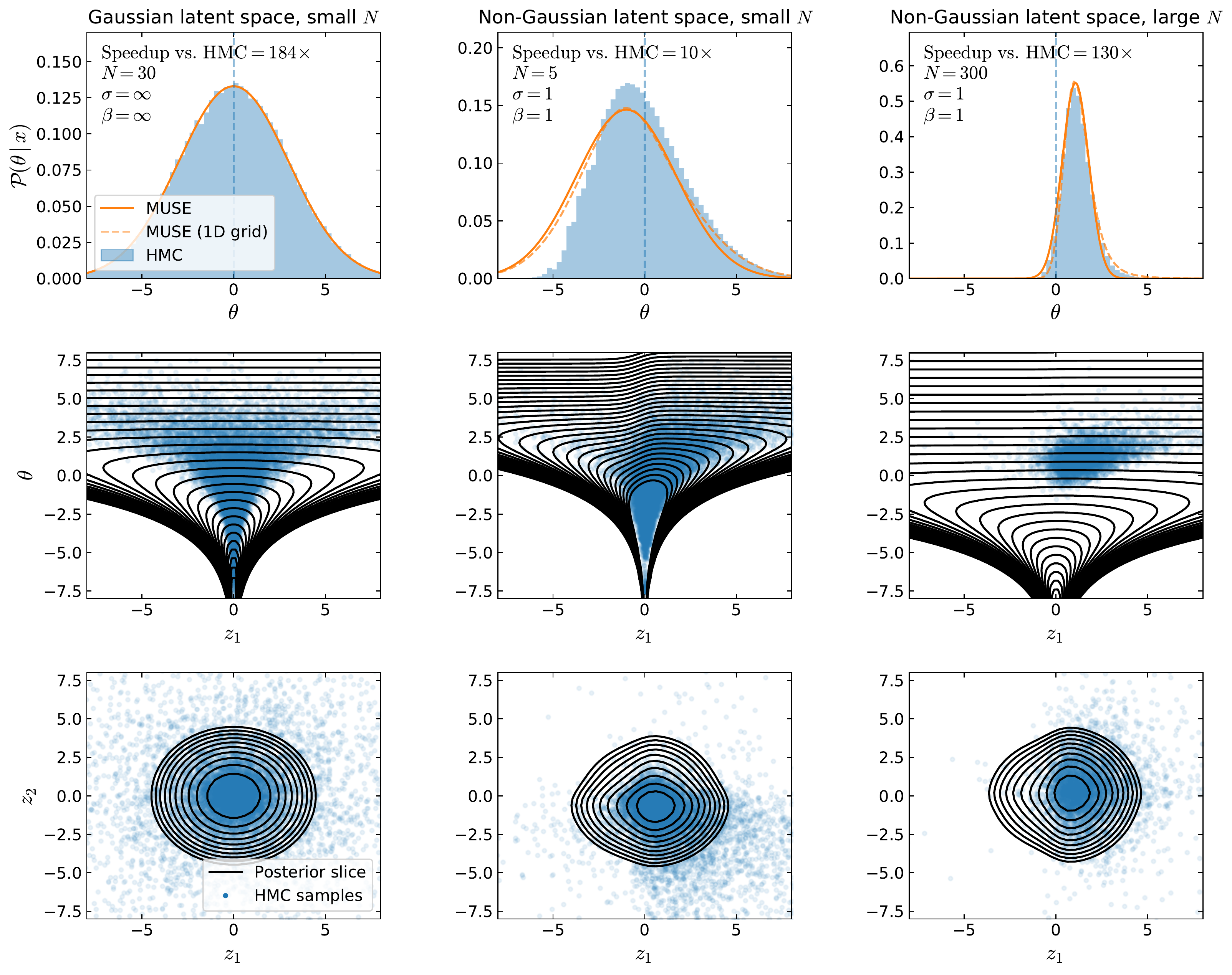}
    \caption{\MUSE applied to the toy ``funnel"-like problem described in Sec.~\ref{sec:toympm}. Each column corresponds to a run using the $N$, $\sigma$, and $\alpha$ parameters indicated in the top panel. The top row compares the \MUSE estimate (solid orange) with the exact posterior determined via HMC (histogram). Also shown (dashed orange) is the result of integrating over the \MUSE score evaluated on a grid, which is possible in 1D. In the middle and bottom rows, the black contours show a slice through the joint likelihood with parameters which are not pictured fixed to zero, and the blue dots show HMC samples and map out the marginal distribution. \MUSE is most useful in cases like the final column where the joint likelihood can be non-Gaussian in any parameter, but the posterior on a parameter of interest has been driven to near-Gaussianity by the central limit theorem. Conversely, the middle column is a failure mode of \MUSE due to low dimensionality and latent non-Gaussianity. In the cases here spanning $N\,{=}\,30-300$ latent dimensions, \MUSE outperforms HMC in terms of the number of posterior gradient evaluations by one to two orders of magnitude.}
    \label{fig:toympm}
\end{figure*}

\section{\MUSE on a toy problem}
\label{sec:toympm}

We now demonstrate the \MUSE algorithm on a toy problem. A scenario which arises in many different contexts in Bayesian hierarchical modeling is the so-called ``funnel problem'' \cite{neal2003}, which exhibits both non-Gaussianity and is particularly difficult to sample with HMC without additional tricks \citep{papaspiliopoulos2007}. The funnel problem is a standard benchmark for inference algorithms, and is a useful problem to build intuition about the \MUSE estimate.  

We also slightly extend the funnel problem for the purposes of demonstration. The extended version embeds the funnel within one additional hierarchical layer, and contains some tunable parameters which can further increase the non-Gaussianity. It is defined by
\begin{align}
    \theta &\sim {\rm Normal}(0, 3) \\ 
    z_i    &\sim {\rm Normal}(0,e^{\theta/2}) \\ 
    \x_i   &\sim {\rm Normal}\big(\beta \tanh(z/\beta), \sigma\big)
\end{align}
The nomenclature is the same as in the previous section: $\theta$ is the parameter of interest, $z$ are the latent space variables, and $\x$ are the observed variables. The tunable parameter $\beta$ can increase the non-Gaussianity of the latent space; in the limit $\beta\rightarrow\infty$, the latent space is Gaussian when conditioned on $x$ and $\theta$, whereas for $\beta\,{\sim}\,1$, non-Gaussianity is induced even just within the latent space conditional. The latent space and data are indexed by $i=1...N$ where $N$ controls the number of data points and hence how closely we approach the asymptotic limit. The parameter $\sigma$ controls ``noise'' in the observations. In the limit $\sigma\,{\rightarrow}\,\infty$ and $\beta\,{\rightarrow}\,\infty$, we recover exactly the standard funnel problem.

As a short aside from the general discussion, we note that funnel-like problems are extremely common in cosmological inference problems since theories generally predict the statistical properties of some field, rather than the field itself. For example, the CMB power spectrum predicts the distribution of fluctuations in the CMB temperature and polarization fields, and the galaxy power spectrum predicts the distribution of fluctuations in the galaxy density field. Inferring either power spectrum amounts to inferring a variance parameter analogous to $\theta$ in the funnel problem, with the field serving as $z$. 

We consider three cases of our toy problem, pictured in the three columns in Fig.~\ref{fig:toympm}, which demonstrate important regimes for the \MUSE estimate. In each case, we compare the \MUSE estimate with the exact posterior computed with the more expensive but exact HMC procedure. In all cases, we set the true value of $\theta$ to zero and generate a sample of $x$ from the forward model, then use \MUSE or HMC to estimate $\mathcal{P}(\theta\,|\,x)$. Because this toy problem features a prior on $\theta$, we are using the Bayesian version of \MUSE described in the previous section. 

We benchmark the two algorithms by comparing the number of joint likelihood gradient evaluations needed to estimate the posterior mean of $\theta$ to similar accuracy. Of course, HMC can always produce the exact posterior mean whereas \MUSE will sometimes be approximate, so this comparison should only be regarded as fair for the problems with near-Gaussian $\theta$ posteriors that are targeted by \MUSE. In the case of HMC, the error on the posterior mean relative to the standard deviation will scale as $1/\sqrt M$ where $M$ is the number of statistically independent samples in the chain (which can be computed from the effective sample size). For \MUSE, the error will scale as $1/\sqrt{M}$ where $M$ is the number of simulations used to compute the expectation value in $\smuse$. We thus compare the number of gradient evaluations needed for one independent sample vs. the number needed for one of the \MUSE simulations throughout the course of the entire iterative procedure. For \MUSE, the computation of $J$ does not add additional cost since the relevant sims are already be computed for the purposes of the estimate itself, and $H$ is usually a subdominant cost since it is only weakly data-dependent, needs very few sims, and can often be speed up with AD. 

For HMC, we use the NUTS algorithm as implemented in \textsc{Turing.jl} \cite{ge2018}. The target acceptance rate is set to 0.65 (the optimal choice for a Gaussian \cite{beskos2010}), except for the first case where it is set to 0.999 to adequately sample the tails of the distribution \cite{modi2021}. For \MUSE, we use our package \textsc{MuseInference.jl}, with a step size $\alpha=0.7$, relative error tolerance on $\theta$ of 1\% of the standard deviation, and an absolute tolerance on the gradient at the MAP solution of $10^{-4}$. We now describe the three cases:

\paragraph*{$(N\,{=}\,30, \, \sigma\,{=}\,\infty, \, \beta\,{=}\,\infty)$}  This case is the standard funnel problem. It features a Gaussian latent space and an $N$ which is too small to achieve the asymptotic guarantees of \MUSE. Nevertheless, \MUSE still recovers the true posterior perfectly because $\smuse$ is always exact for Gaussian latent spaces. The black contours in the bottom row show a slice through the $(z_1, z_2)$ posterior with all other $z_i$ and $\theta$ set to zero, demonstrating the latent Gaussianity. Note that the $(\theta, z_i)$ posterior can still be (highly) non-Gaussian, as shown by the black contours in the middle row, which instead exhibit the ``funnel'' shape which is namesake of the problem. The top panel shows the near-perfect overlap between a histogram of the HMC samples and the \MUSE posterior estimate. Blue dots in the lower two panels also show HMC samples for reference (note the difference between the blue points which sample the posterior {\it marginalized} over the other variables, vs. the black contours which fix these other variables to zero). In terms of benchmarks, the funnel problem is nearly trivial for \MUSE to solve. Because here the data is uninformative, the MAP solution is $\zMAP=0$ for any $x$ or $\theta$, which is achieved with only a few posterior gradient evaluations (we do not put in the solution by hand, instead letting the optimizer proceed as usual). The \MUSE posterior is then simply given by the Gaussian expansion of the prior, which gives the exact answer here. Conversely, HMC struggles on this problem due to the wide range of scales which must be traversed by the symplectic integrator in the mouth vs. neck of the funnel. Overall, we find \MUSE requires 184 times fewer gradient evaluations than HMC on this case. We stress that while this toy problem is certainly expected to highlight the benefits of \MUSE, it can be regarded as fair since no special information was input to \MUSE beyond the standard generic procedure described in the previous section. While there are known ways to greatly improve HMC performance on the funnel problem \cite{papaspiliopoulos2007,millea2020b}, these require manual intervention, whereas \MUSE will automatically perform well on funnel-like problems, or subspaces of more complex problems which exhibit funnel-like behavior, even if these subspaces are not known a priori.
    
\paragraph*{$(N\,{=}\,5, \, \sigma\,{=}\,1, \, \beta\,{=}\,1)$} This case sets $\sigma=\beta=1$ and reduces the dimensionality to $N=5$. We are now even further from the asymptotic limit, and the latent space is no longer Gaussian. This is the main failure of mode of \MUSE. Here, there are no guarantees that the inference is unbiased or that it has the correct variance. Indeed, in the top center panel we see neither are correct when compared to the HMC posterior. If for some real-world problem one suspects that this failure mode is being hit, one diagnostic is to run \MUSE on a suite of simulations with a known input truth and to check if, on average, the truth is recovered and if the empirical covariance matches Eqn.~\eqref{eq:bayesmuse}.
    
\paragraph*{$(N\,{=}\,300, \, \sigma\,{=}\,1, \, \beta\,{=}\,1)$} This is like the previous case except we have increased the dimensionality to $N\,{=}\,300$. This is large enough that the asymptotic guarantees of \MUSE kick in, and the top-right panel of Fig.~\ref{fig:toympm} shows that we again recover the HMC posterior near-perfectly. In doing so, we find that \MUSE uses 130 times fewer posterior gradient evaluations than HMC. This case is the closest to where we envisage \MUSE is most useful in the real-world: high-dimensional problems where the joint posterior need not be Gaussian in any variable, but where the marginal posterior on parameters of interest is asymptotically driven to near-Gaussianity. The CMB lensing problem which we will discuss in the next section is most similar to this third example, featuring a latent space with dimensionality $N\,{\sim}\,10^5\,{-}\,10^7$. 

For this last case, we also show in Fig.~\ref{fig:toympm} the result of computing $\smuse$ on a grid of $\theta$ values and integrating the result to produce an exact \MUSE posterior approximation (possible only in 1D). This allows us to visualize the quality of the \MUSE approximation separately from its Gaussianization around the peak, and note that it tracks the true posterior extremely well. Although computationally costier, this can be a valid way to deal with cases where there is only a single parameter of interest. 

\section{\MUSE CMB lensing}
\label{sec:lensingposterior}

\subsection{The CMB lensing problem}

We now describe the CMB lensing problem, which we use to demonstrate \MUSE in a challenging real-world scenario, and for which \MUSE provides a novel solution. The goal of this analysis is to estimate the power spectrum of the CMB lensing potential and the power spectrum of the unlensed $E$-mode polarization given noisy lensed CMB data. We ignore CMB temperature because this leads to smaller differences between Bayesian and QE methods, and we do not estimate $B$-mode bandpowers since including these is qualitatively the same as the $E$-modes in terms of \MUSE ($B$-mode polarization maps do enter the algorithm, just with a theory spectrum which is assumed perfectly known). It is straightforward to include either of these components if desired. The lensing problem can be summarized as
\begin{align}
    (\AEb, \Aphib) &\sim {\rm Uniform}\big(0,\infty\big) \\
    f &\sim \Cf[\AEb] \\
    \phi &\sim \Cphi[\Aphib] \\
    \x &\sim {\rm Normal}\big(\mathbb{A} \,\Len[\phi]f, \, \Cn\big),
\end{align}
with the corresponding joint likelihood distribution
\begin{widetext}
\begin{align}
\label{eq:jointposterior}
\mathcal{P}(\x,\underbrace{f,\phi}_{z}\,|\,\underbrace{\AEb,\Aphib}_{\theta}) &\propto 
\frac{\exp\left\{ -\cfrac{\big(\x - \op A\, \Len[\phi] \, f \big)^2}{2 \,\Cn} \right\}}{\det \Cn^{\nicefrac{1}{2}}} \;
\frac{\exp\left\{ -\cfrac{f^2}{2\,\Cf[\AEb]} \right\}}{\det  \Cf[\AEb]^{\nicefrac{1}{2}}} \;
\frac{\exp\left\{ -\cfrac{\phi^2}{2\,\Cphi[\Aphib]} \right\}}{\det  \Cphi[\Aphib]^{\nicefrac{1}{2}}} \mathcal{P}_{S}(\phi)
\end{align}
\end{widetext}
where
\begin{itemize}
    \item $\op A \equiv  \op{M} \cdot \op{T} \cdot \op{B}$ is a linear operator containing the instrumental beam, $\op{B}$, transfer function, $\op{T}$, and any masking, $\op{M}$
    \item $f$ is the map of the unlensed CMB polarization
    \item $\phi$ is the map of gravitational lensing potential
    \item $\x$ is the data
    \item $\Len[\phi]$ is a linear operator which lenses a map (and whose dependence on $\phi$ is non-linear)
    \item $\Cn$, $\Cf$, and $\Cphi$ denote the covariances for the noise, unlensed CMB, and $\phi$, respectively
    \item $\AEb$ and $\Aphib$ are bandpower amplitudes which control the covariances $\Cf$ and $\Cphi$, respectively. Specifically, they scale the isotropic CMB polarization and lensing power spectra as
    \begin{align}
        C_\ell^{\phi\phi} &= \left[1+\sum_b (\Aphib-1) \, W^{\phi}_{b\ell} \right] C_\ell^{\phi\phi,\rm fid} \\
        C_\ell^{EE} &= \left[1+\sum_b (\AEb-1) \, W^{E}_{b\ell} \right] C_\ell^{EE,\rm fid},
    \end{align}
    where bandpower window functions $W_{b\ell}\,{=}\,1$ if $\ell$ falls within bin $b$, and is zero otherwise. We chose to estimate amplitudes relative to a fiducial model rather than the power spectra themselves for simplicity and without loss of generality. The form of $W$ is arbitrary and chosen for simplicity, and should not matter as long as the spectra do not vary significantly across the bin. Here we use a typical binning used in previous SPT analyses which features $\Delta\ell\,{=}\,50$ for $E$ and 10 logarithmically spaced bins for $\phi$. 
    \item $\mathcal{P}_{S}(\phi)$ is a ``super-sample'' prior on $\phi$ which we discuss below.
\end{itemize}
As indicated in Eqn.~\eqref{eq:jointposterior}, $\Aphib$ and $\AEb$ form the $\theta$ parameters which wish to infer, and the maps of the unlensed CMB and of the lensing potential, $f$ and $\phi$, form the latent space, $z$. Depending on the pixelization and area of sky observed, these maps can easily contain more than a million pixels, and although the likelihood is Gaussian in $f$, it is a non-Gaussian function of $\phi$. This high-dimensionality and non-Gaussianity makes the CMB lensing problem an excellent test-bed for \MUSE.

The joint likelihood shown in Eqn.~\eqref{eq:jointposterior} is exactly as described in previous works which have attempted to maximize or sample this distribution \cite{anderes2015,millea2019,millea2020b}. Here, we make one additional and simple change which we find is very helpful in making the problem more amenable to MUSE estimation. As discussed in \cite{millea2019}, in cases where a pixel mask is present (which is the case for any real analysis), the MAP estimate when maximizing jointly over both $f$ and $\phi$ incurs a contribution from a ``mean-field.'' The mean-field approximately manifests as an additive offset to the magnification, $\kappa\,{\equiv}\,-\nabla^2\phi/2$, across unmasked pixels. It arises because the mean magnification is otherwise very unconstrained due to aliasing from the mask. The presence of the mean-field can imprint a bias in the MUSE estimate because these unconstrained modes are then aliased into the bandpowers of interest, but do not benefit from guarantees of asymptotic unbiasedness present for other modes. A simple resolution which we have found is to impose a prior
\begin{align}
    \label{eq:supersample}
    \mathcal{P}_S(\phi) = \exp\left(-\frac{\big(\langle -\nabla^2\phi/2 \rangle_{\rm unmasked\;pixels}\big)^2}{2\sigma_\kappa^2}\right),
\end{align}
which recenters the mean $\kappa$ in the joint MAP estimate to near zero.  The prior almost perfectly removes the mean-field from the joint MAP estimate, and otherwise does not require any modification to the generic \MUSE procedure (it is simply an additional contribution to the $\mathcal{P}(z\,|\,\theta)$ term in Eqn.~\ref{eq:likelihood}). The prior can be motivated physically by noting that its need arises from the data failing to constrain modes in the $\kappa$ map that are larger than the observed field. However, theoretically we know these ``super-sample'' modes should be near zero. Alternatively, for a given region of sky, full-sky {\it Planck} observations may give a good estimate of what the magnification actually is. It is thus valid to impose this as a prior, with the added benefit that at the same time it remedies possible biases in the \MUSE estimate from this effect.

Finally, we note that instead of the joint distribution in Eqn.~\eqref{eq:jointposterior}, it would in theory be possible to analytically marginalize over $f$, then perform \MUSE on just the remaining part of the distribution, $\mathcal{P}(\phi,\theta\,|\,\x)$. This form of the posterior was explored by \cite{hirata2003a,carron2017}, with some further discussion in \cite{millea2019}. While it may seem advantageous to perform as much of the marginalization as possible analytically, in terms of speed it would actually be a detriment because the subsequent gradients and MAP estimates of the marginal distribution are far more computationally costly than of the joint. It is instead much faster to work with the original joint distribution and simply allow the integral over $f$ to be performed implicitly as part of the \MUSE procedure. Because \MUSE is exact for Gaussian conditional slices (e.g. Appendix~\ref{app:exactmpm}), this will not introduce any extra approximations. One can in fact view the marginal MAP $\phi$ estimate of \cite{carron2017} as exactly equivalent to running \MUSE with $\theta=\phi$, $z=f$, and fixed cosmological parameters.

\begin{figure*}
    \includegraphics[width=\textwidth]{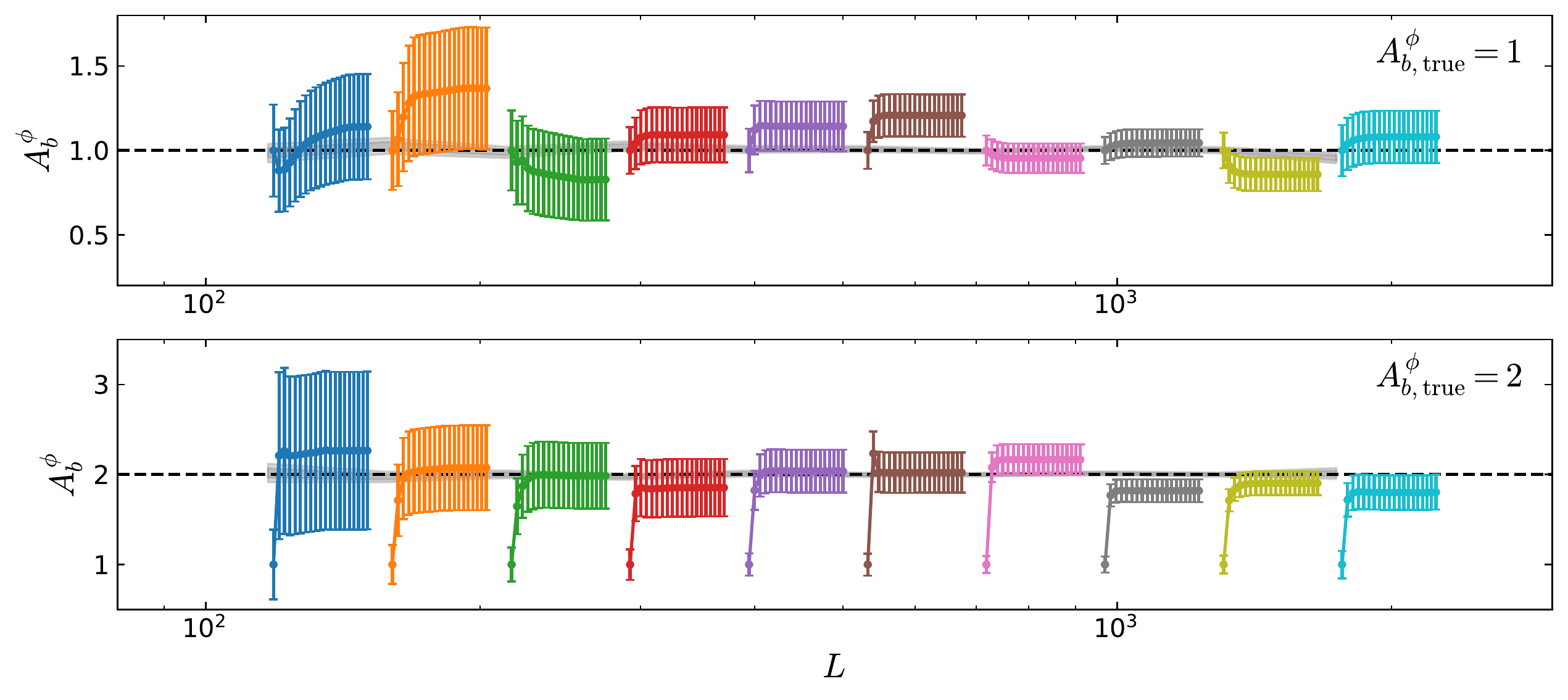}
    \caption{Examples of the \MUSE estimate of the binned bandpower parameters, $\Aphib$, for a simulated \configSMALL dataset, including masking. Each color represents a different bandpower, and within each bandpower, each iterative \MUSE step is arbitrarily offset in $\ell$ for clarity. In the top panel, the true value is $\Aphib=1$ (for all $b$), and the bottom panel the true value is $\Aphib=2$. Both cases start from an initial guess of $\Aphib=1$. Each panel also includes a gray band showing the 1\,$\sigma$ bound on the bias in each bandpower, computed from an average over many simulated realizations (the bias is asymptotically zero, and demonstrated above to be sufficiently near-zero as well for the finite number of modes constraining each bandpower here).}
    \label{fig:phibandpowerssmall}
\end{figure*}

\begin{figure}
    \includegraphics[width=0.9\columnwidth]{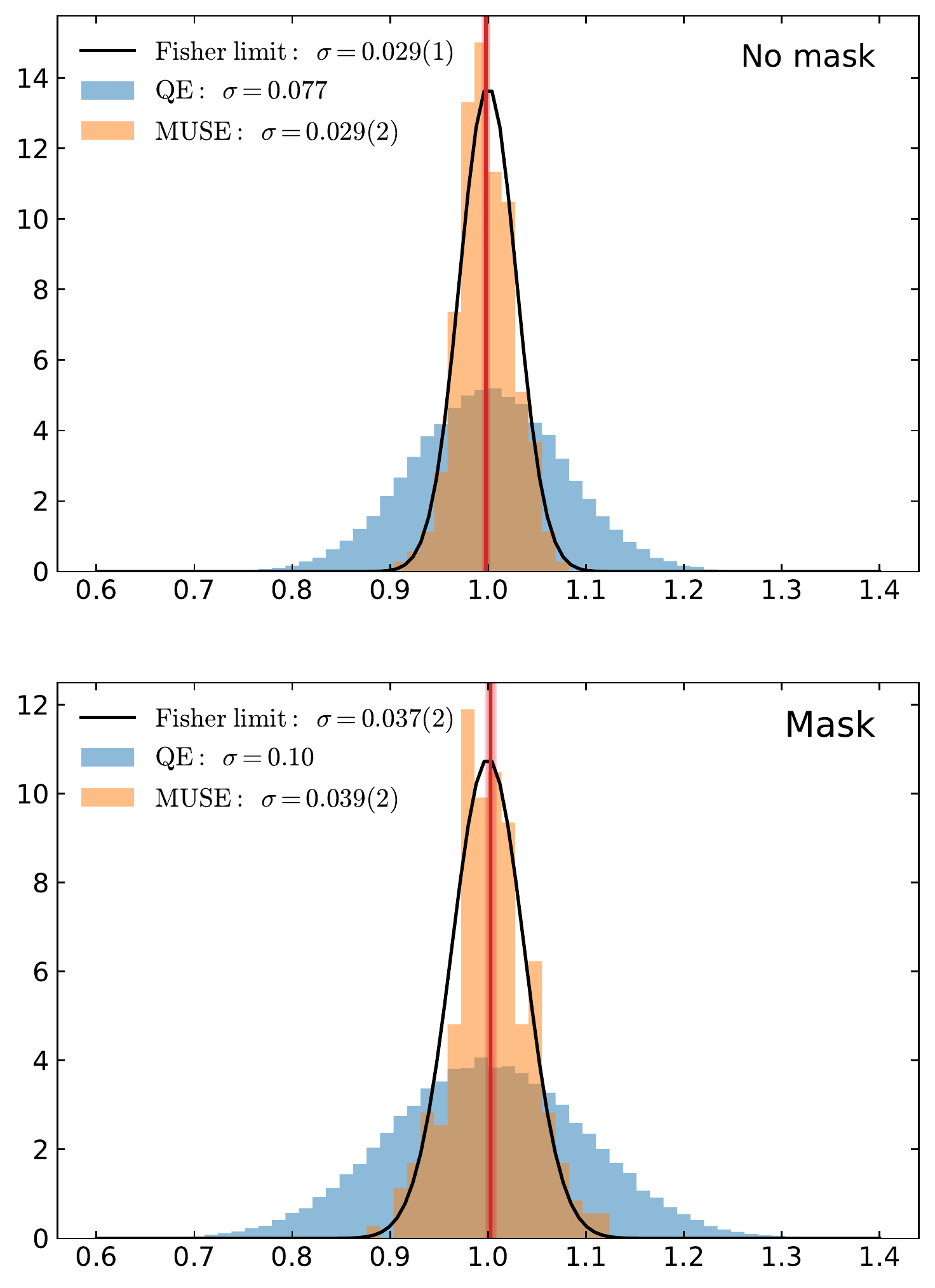}
    \caption{Empirical scatter from simulations in $\Aphi$ estimates computed from \MUSE lensing bandpowers (orange bars) or quadratic estimate lensing bandpowers (blue bars), as compared to the exact Fisher information computed from MCMC chains (black line). The numbers in parenthesis in the legend give the sampling error on the final digit of these standard deviations due to the finite number of simulations which entered these calculations. The red vertical band gives the mean of the \MUSE estimates and the 1 and 2 $\sigma$ standard error on the mean. This plot demonstrates the \MUSE result is consistent with being unbiased and saturates the Cram\'er-Rao bound (i.e. is effectively optimal) to better than ${\sim}\,10$\% even in the presence of masking.}
    \label{fig:hists}
\end{figure}

{
\renewcommand{\arraystretch}{1.2}
\setlength{\tabcolsep}{8pt}
\begin{table}
    \begin{tabular}{l|c|c}
        \toprule
        Simulation & \configSMALL & \configBIG  \\
        \colrule
        Map size                               & 256$\times$256          & 1024$\times$2048       \\
        Pixel width                            & 3\,arcmin               & 2.1\,arcmin            \\
        Total area                             & $\sim$150\,deg$^2$      & $\sim$1500\,deg$^2$    \\
        Noise level in $T$                     & $1\;\mu$K-arcmin        & $2.3\;\mu$K-arcmin     \\
        $(\ell_{\rm knee}, \alpha_{\rm knee})$ & (100,3)                 & (100,3)                \\
        Beam FWHM                              & 2\,arcmin               & 1\,arcmin              \\
        Fourier masking $(\op{K})$             & $2<\ell<3300$           & $2<\ell<5000$          \\
        Pixel masking $(\op{M})$               & varying                 & SPT-3G                 \\
        Fiducial $r$                           & $r=10^{-1}$             & $r=10^{-1}$            \\
        \toprule \MUSE estimation && \\
        \colrule
        Estimated $\theta$                     & $\Aphib$ and $\AEb$     & $\Aphib$ and $\AEb$ \\
        \# of simulations                      & 100                     & 100                    \\
        Step size                              & 0.7                     & 0.7                    \\
        \# of steps                            & 10                      & 10                     \\
        \MUSE runtime                          & 5\,minutes              & 60\,minutes            \\
        HMC runtime                            & 5\,hours                & $\sim$\,week (est.)       \\
        \colrule
    \end{tabular}
    \caption{Simulation parameters and \MUSE solution parameters for the different configurations of simulated data used in this work. Listed runtimes are wall-time using four Tesla A100 GPUs. HMC runtime refers to the time to run an MCMC chain sampling just a single bandpower parameter whereas the \MUSE runtime estimates the full set of bandpowers.}
    \label{table:configurations}
\end{table}
}

\subsection{Lensing bandpowers}

As a first check, we run \MUSE on a suite of simulated 256$\times$256 pixel $150$\,deg$^2$ patches of sky with noise levels similar to those planned for CMB-S4. Although a smaller sky area than the CMB-S4 footprint, we do not expect the accuracy of the \MUSE estimate to depend significantly on the size of the observed sky area, since lensing is a fairly local operation (lensing deflections are coherent across only a few degrees). Limiting the size of the dataset allows us to run a large number of simulations and very accurately quantify the properties of the \MUSE estimate. To start, we will fix $\AEb\,{=}\,1$ and estimate only $\Aphib$. In the next subsection we will demonstrate estimating both simultaneously. We also consider two versions of this case, one without a pixel mask, and another with a $1^\circ$ mask around the edges of the field. We refer to this dataset as the \configSMALL data, and exact simulation parameters are given in Table~\ref{table:configurations}. 

We run \MUSE as described in Sec.~\ref{sec:mpm}. Here, the $\zMAP$ which appears in the procedure is the best-fit $(f,\phi)$ at fixed $\Aphib$:
\begin{align}
    (\hat f_J, \hat \phi_J) = \underset{f,\phi}{\rm argmax} \; \log \mathcal{P}(\x,f,\phi\,|\,\Aphib),
\end{align}
We follow previous CMB lensing literature in denoting the joint MAP as $(\hat f_J, \hat \phi_J)$. The maximization is performed iteratively using the coordinate descent algorithm presented in \cite{millea2019}. We then solve the \MUSE equation to obtain the estimate of the bandpower mean, $\hAphib$, which solves
\begin{align}
    \smuse(\hAphib) = 0.
\end{align}
For convenience we use AD to compute the gradients of Eqn.~\eqref{eq:jointposterior} with respect to $\Aphib$, although the analytic gradient is simple to derive as well. We use 100 simulations to compute the expectation value in $\smuse$, corresponding to a Monte Carlo error of ${\lesssim}\,0.1\sigma$. We find ${\sim}\,15$ Broyden iterations are sufficient for the bandpowers to converge. 

We note that previous works based on HMC sampling found it necessary to work with a reparameterized form of Eqn.~\eqref{eq:jointposterior} which decorrelated $f$ and $\phi$, and removed funnel-like correlation between them and $\Aphib$ (this was denoted the ``mixed posterior''). Although we still use mixing for speeding up the computation of the joint MAP estimate, we highlight that no such reparameterization is needed for the \MUSE estimate itself, which performs identically whether or not we use mixing. This is due to the excellent performance of \MUSE on funnel-like problems which was highlighted in the previous section.

The majority of the runtime is spent computing the joint MAP for the data and for the simulations in each \MUSE iteration. As mentioned earlier, it is crucial for performance to use the joint MAP estimate from the previous iteration as a starting point for the joint MAP estimate at the next iteration, both for the data and for the suite of same-seeded simulations. This reduces the overall runtime by almost an order of magnitude, since in later iterations, $\theta$ is not changing much and the starting points are extremely close to the final estimate. The entire procedure runs in about 5 minutes on four Tesla A100 GPU for this problem size.

In Fig.~\ref{fig:phibandpowerssmall}, we plot the resulting $\hAphib$ estimates after each Broyden iteration for a simulated masked \configSMALL dataset. This shows a typical evolution of the bandpower parameters as the \MUSE solution is iteratively obtained. Error bars have been plotted at each step for demonstration, although in practice they would only be calculated for the last step. The final result scatters around the input fiducial model, which is a simple sanity check that the estimate is unbiased. More quantitatively, we check the bias is sufficiently small using simulations. Although the bias is asymptotically zero, it may be non-zero for any finite data vector, and scales as $1/\sqrt{N}$ as demonstrated in Sec.~\ref{sec:freqmuse}. Since $N$ is non-trivial to estimate, here we empirically determine the bias directly. The gray band in Fig.~\ref{fig:phibandpowerssmall} shows the mean over 512 simulated \MUSE analyses and its 1\,$\sigma$ standard error. We find no evidence for any bias at the level of 0.05\,$\sigma$ which is afforded by this number of simulations. The bottom panel of Fig~\ref{fig:phibandpowerssmall} shows a case where the fiducial $\Aphib\,{=}\,2$, but the initial starting guess for the \MUSE estimate is $\Aphib\,{=}\,1$. The colored error bars show how the estimate iteratively moves towards the higher value of $\Aphib$. The gray band similarly shows an average over 512 realizations, and demonstrates the bias is consistent with zero for this alternate fiducial model as well. We conclude that down to the noise level of CMB-S4, for similarly wide bandpowers, and for sky area of 150\,deg$^2$ or larger (the number of modes would grow with larger sky area, reducing the estimator bias), the \MUSE lensing estimate is effectively unbiased.

Next, we consider the optimality of the estimate. Because the bias is effectively zero, the covariance must satisfy the Cram\'er-Rao bound. Writing $\Sigma\,{\equiv}\,\Sigma^{\rm MUSE}$ for brevity, the bound states that for all $\theta$,
\begin{align}
    \Sigma_{bb^\prime} - \mathcal{F}^{-1}_{bb^\prime} \geq 0,
\end{align}
where the inequality represents that the left-hand side must be a positive semi-definite matrix, and for an optimal estimate this becomes an equality. At present, there is no exact way to calculate $\mathcal{F}_{bb^\prime}$ for the full set of bandpowers.\footnote{Sometimes the iterative power spectrum forecasts of \cite{smith2012} are taken as a benchmark of optimality. While these are extremely useful, we note that they are only approximations rather than formal calculations of Fisher information, and it is unknown how they perform in the presence of effects like masking which we wish to check here.} However, we can obtain $\mathcal{F}$ for an overall amplitude parameter which is derived as a minimum variance combination of $\hAphib$
\begin{align}
    \hAphi \equiv w_b \hAphib
\end{align}
where
\begin{align}
    w_b = \frac{\sum_{b^\prime} C_{bb^\prime}^{-1}}{\sum_{b,b^\prime} C_{bb^\prime}^{-1}}
\end{align}
The posterior distribution for this same parameter, $\mathcal{P}(\Aphi\,|\,\x)$, can be computed via MCMC using the method described in \cite{millea2020b}. If a flat prior on $\Aphi$ is assumed, this also equals the likelihood, $\mathcal{P}(\x\,|\,\Aphi)$. With the likelihood obtained in this manner, the Fisher information can be computed as
\begin{align}
    \mathcal{F} = \left \langle \frac{d^2}{d\Aphi^2} \log \mathcal{P}(\x\,|\,\Aphi) \right \rangle_{\x\sim\mathcal{P}(\x\,|\,\Aphi)}
\end{align}
by explicitly averaging over the curvature of the log-likelihood from several chains on different simulated data. This can then be compared with variance of the $\hAphi$ estimator computed from simulations.\footnote{In theory, one could use HMC sampling to infer the full bandpower posterior, $\mathcal{P}(\Aphib\,|\,d)$, and estimate the Fisher information matrix in this way. This may be possible in practice, but has yet to be demonstrated, and was not attempted here due to its additional computational cost.}

Fig.~\ref{fig:hists} shows \MUSE $\hAphi$ estimates from a suite of 256 simulations. For comparison, results from the quadratic estimator (QE) are also shown for a suite of $10^5$ simulations (these are computationally inexpensive to compute). The black curve is a Gaussian with standard deviation given by $\mathcal{F}^{-1/2}$, computed via MCMC chains as just described. Because the scatter in the curvature of the log-likelihood is small (put another way, the error bars on $\Aphi$ from each chain are only mildly realization-dependent), only a small number of chains are needed to compute the Fisher information to within acceptable Monte Carlo error, here only ten chains. 

Independent of whether we apply pixel masking or not (top and bottom panels in Fig.~\ref{fig:hists}), we find the \MUSE results are consistent with saturating the Cram\'er-Rao bound. This is encouraging, because the pixel mask induces additional non-Gaussianity of the latent space posterior (beyond that present due simply to lensing), which one could have been suspected to cause sub-optimality in the \MUSE estimator. Evidently, however, these are not large enough to significantly degrade its optimality.

Because $\hAphi$ saturates the Cram\'er-Rao bound, an even stronger statement can be derived, mainly that $\hAphib$ also saturates the bound. Intuitively, this is because Fisher information is always positive, so if the Fisher information for any function derived from the bandpowers is maximized, the Fisher information in each individual bandpower must also be maximized. More rigorously, consider the covariance and Fisher information for $\Aphi$ and $\Aphib$, 
\begin{align}
    \Sigma &\equiv \sum_{b,b^\prime} w_b \Sigma_{bb^\prime} w_{b^\prime} \\
    \mathcal{F}^{-1} &\equiv \sum_{b,b^\prime} w_b \mathcal{F}^{-1}_{bb^\prime} w_{b^\prime}
\end{align}
Since we have empirically verified that $\Sigma\,{=}\,\mathcal{F}^{-1}$ (up to Monte Carlo error), we have that
\begin{align}
    \sum_{b,b^\prime} w_b \big( \Sigma_{bb^\prime} - \mathcal{F}^{-1}_{bb^\prime}\big) w_{b^\prime} = 0 \\
    \sum_{b} v_b^2 \big( \tilde{\Sigma}_{b} - \tilde{\mathcal{F}}^{-1}_{b}\big) = 0
    \label{eq:diagsum}
\end{align}
where in the second equation we have simultaneously diagonalized the covariance and Fisher matrix (possible because both are positive definite), yielding new weights, $v$ and diagonal entries, $\tilde{\Sigma}_b$ and $\tilde{\mathcal{F}}_b$. Since the Cram\'er-Rao bound guarantees that $\tilde{\Sigma}_{b} - \tilde{\mathcal{F}}^{-1}_{b}$ is a positive number, it follows that every term in the sum in Eqn.~\eqref{eq:diagsum} must be zero individually, and hence that $\Sigma_{bb^\prime} = \mathcal{F}^{-1}_{bb^\prime}$.

\begin{figure*}
    \centering
    \includegraphics[width=\textwidth]{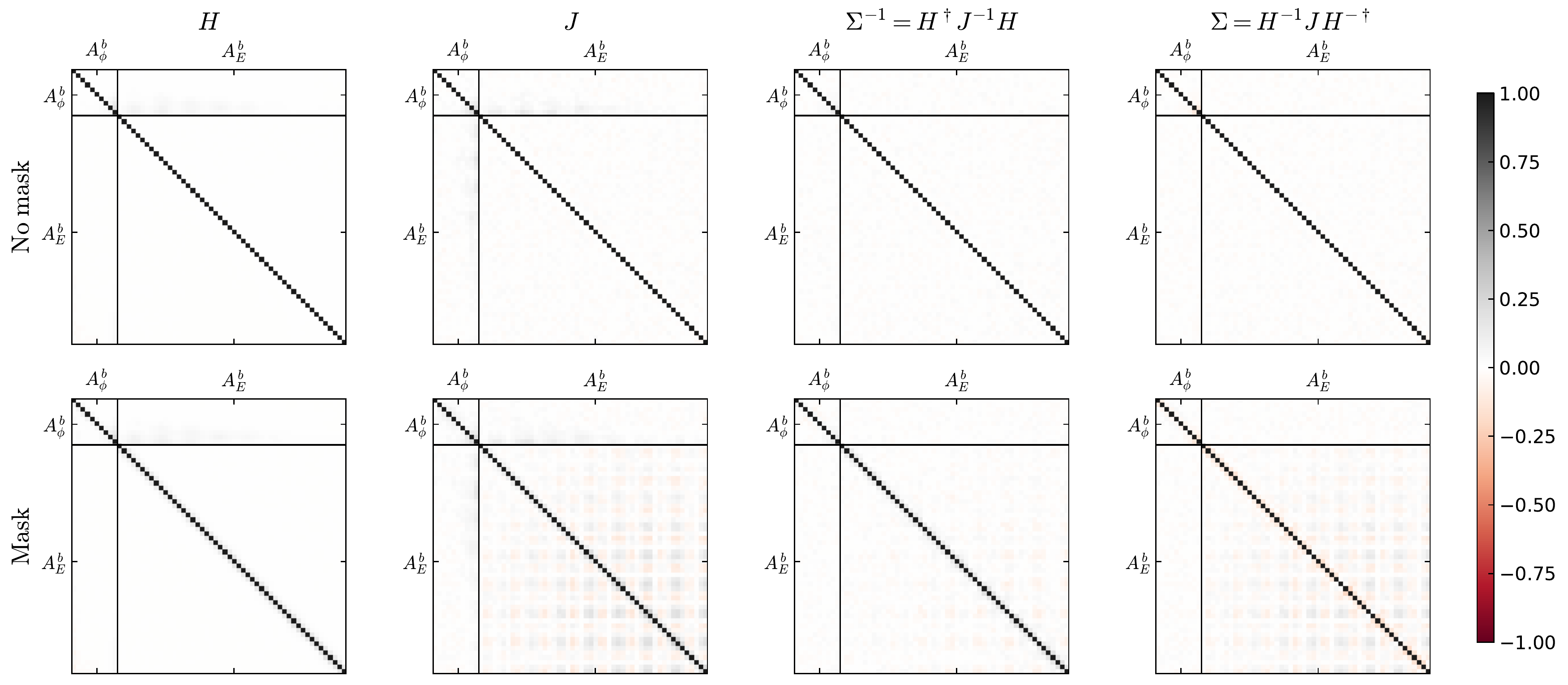}
    \caption{The $J$ and $H$ matrices (defined in Eqs.~\ref{eq:defJ} and \ref{eq:defH}), and the bandpower covariance and its inverse, $\Sigma$ and $\Sigma^{-1}$. The \configSMALL configuration is assumed, and bottom and top rows correspond to with and without masking, respectively. Note the asymmetry visible in $H$, demonstrating that $\smuse$ is non-conservative for the lensing problem. Additionally, we show in the text that the checkerboard pattern in the far off-diagonals does not impact cosmological parameter inferences at more than a few percent $\sigma$.}
    \label{fig:H_J_F}
\end{figure*}

\begin{figure}
    \centering
    \includegraphics[width=\columnwidth]{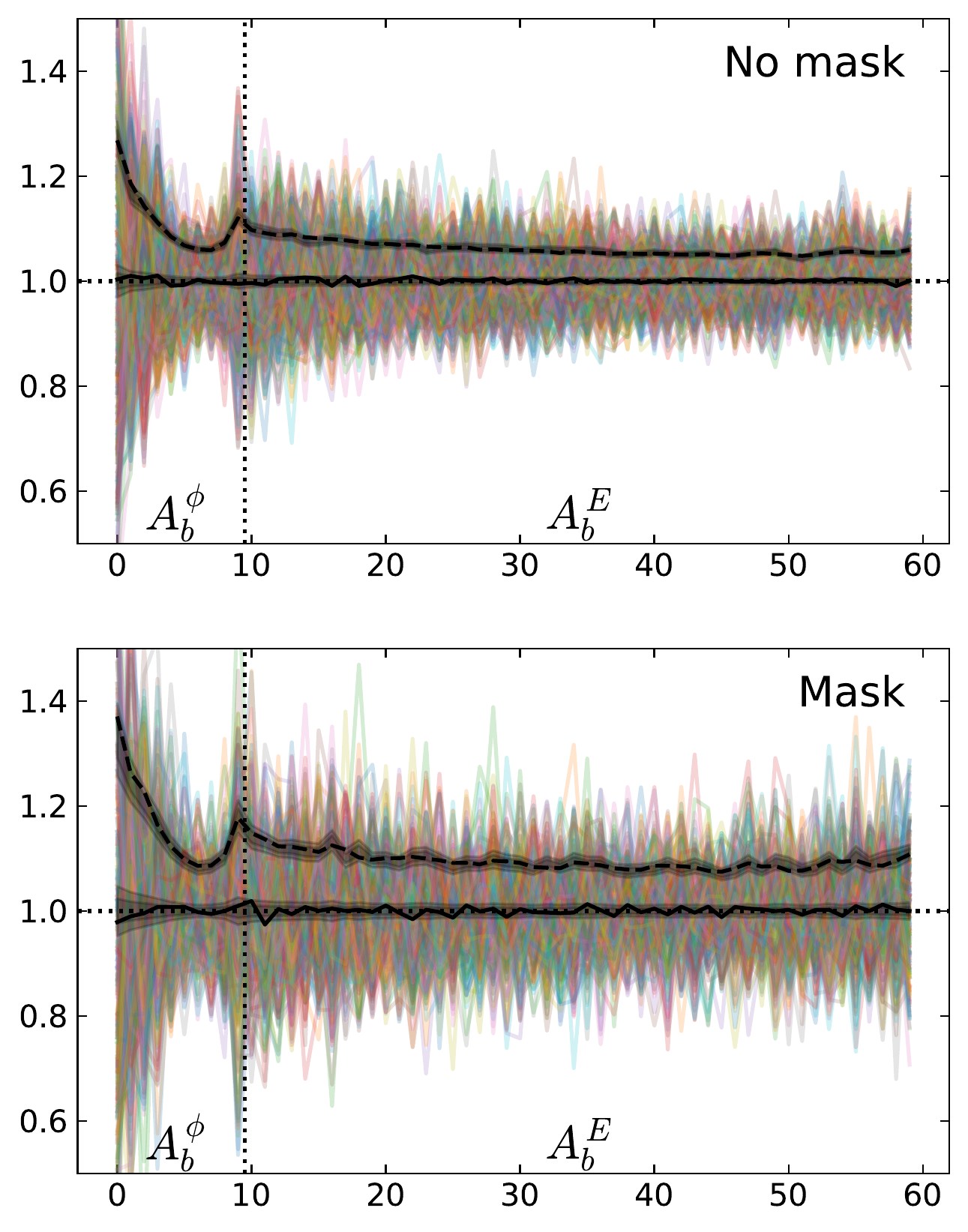}
    \caption{Colored lines show \MUSE estimates of $(\Aphib,\AEb)$ for a large suite of simulations. The \configSMALL configuration is assumed, and bottom and top panels correspond to with and without masking, respectively. The black line around 1 is the mean over all simulations, and the black band around this is standard error on the mean, confirming that \MUSE provides an unbiased estimate of lensing bandpowers joint with delensed $E$ modes. The black dashed lines are the bandpower errors computed from $\surd{\rm diag}(\Sigma^{\rm MUSE})$, and the black bands around this are the empirical scatter and its standard error, demonstrating that the \MUSE covariance prescription accurately reproduces the observed scatter.}
    \label{fig:bias_variance.pdf}
\end{figure}

\subsection{Joint lensing and delensed bandpower estimates}

Next, we consider simultaneously estimating the lensing potential power spectrum and the unlensed $E$ mode bandpowers. Doing so is nearly identical to the previous section except that now $\AEb$ is an estimated parameter in addition to $\Aphib$. This means the MAP estimates that are part of \MUSE now take place at varying values of $\AEb$ as well, 
\begin{align}
    \hat f_J, \hat \phi_J \equiv \underset{f,\phi}{\rm argmax} \; \log \mathcal{P}(\x,f,\phi\,|\,\Aphib,\AEb),
\end{align}
and the estimate is the solution jointly over both sets of bandpowers:
\begin{align}
    \smuse(\hAphib,\hAEb) = 0
\end{align}

We begin by showing in Fig.~\ref{fig:H_J_F} the $H$, $J$, $\Sigma$ and $\Sigma^{-1}$ matrices computed via Eqns.~(\ref{eq:JH}-\ref{eq:asympcov}) for the joint $(\Aphib, \AEb)$ parameter space, with masked and unmasked cases given in the bottom and top rows. These have been computed using 2048 simulations for the $J$ matrix and 64 simulations for the $H$ matrix (the latter which we find is only very weakly realization-dependent). Determining the joint covariance, $\Sigma$, has been a challenging problem for the field even when only the quadratic estimate is used to estimate the lensing potential \cite{schmittfull2013,peloton2017,han2020}. Solutions when using more optimal estimates of the lensing potential have thus far been developed only under simple forecasting assumptions \cite{green2016,hotinli2021}. The result here shows this is now also feasible for a map-level procedure which accounts for masking. 

In the top row (unmasked case) we see there are very few significantly non-zero entries in any of the off-diagonals for any of the matrices. The $H$ matrix does have some small non-zero entries in the upper $\Aphib\times\AEb$ block. As per Eqn.~\eqref{eq:defH}, this corresponds to having injected power into the unlensed signal in some particular $E$ mode bandpower, and the gradient at the MAP having responded instead with a change to a $\phi$ bandpower. This is empirical proof that $\smuse$ is non-conservative (due to the lensing-induced posterior non-Gaussianity), as otherwise the $H$ matrix would be symmetric. In the bottom row (masked case) we additionally see two features: 1) a negative cross-covariance between neighboring $E$-mode bandpowers, and 2) a ``checkerboard'' pattern even for distant $E$-mode bandpower bins. The former effect is typical of induced mode coupling due to the mask. The latter effect is also expected and has been noted by \cite{green2016,hotinli2021}. It arises because the biggest effect of lensing on the $E$-mode power spectrum is a smoothing of the peaks, and this effect is sourced mainly by lensing modes near the peak of the lensing potential power spectrum ($L\sim100$). Thus, depending on if these lensing modes fluctuate high or low, the entire $E$-mode spectrum will be over or under smoothed with respect to the mean theory expectation. This then induces these correlations across very distant neighbors.

Although with \MUSE it is entirely possible to quantify these distant off-diagonal correlations, it is of interest to what extent they actually impact parameter inference and whether they can simply be ignored. To do so, we propagate the \MUSE bandpower covariance, $\Sigma_{bb^\prime}$, to the Fisher information matrix on cosmological parameters, $\mathcal{F}_{\alpha\beta}$, via
\begin{align}
    \mathcal{F}_{\alpha\beta} = \frac{d\log C_b}{d\alpha} \Sigma^{-1}_{bb^\prime} \frac{d\log C_{b^\prime}}{d\beta}
\end{align}
where $\alpha,\beta$ represent cosmological parameters from a standard set $\{\omega_{\rm b}, \omega_{\rm m}, \Sigma m_\nu, \theta_s, A_{\rm s}, n_{\rm s}\}$, and the $\log$ appears because in our definition, $\Sigma$ is the covariance matrix for bandpower amplitudes, rather than bandpowers themselves. We then invert $\mathcal{F}_{\alpha\beta}$ for the case where the full $\Sigma_{bb^\prime}$ is used vs. where only the first off-diagonals are kept, and examine the square root of the diagonal entries (i.e. the forecasted standard deviation error on parameters). We find that no parameter error changes by more than 1\%. This can also be interpreted as that any potential biases due to ignoring these entries would be less than 1\% of the parameter error. We thus conclude that for the \configSMALL configuration, it would be safe to ignore these entries. Note that this includes ignoring the $\Aphib\times\AEb$ cross-covariance entirely. Additionally, because lensing is a relatively local operation, this statement is not expected to depend significantly on the size of the field, thus likely holds for the larger sky fractions targeted by CMB-S4. We also check whether keeping only the diagonal entries of the covariance could be a sufficiently good approximation. Here, we find a 10\% change in parameter error and corresponding 10\% possible bias. This is likely too large to be acceptable, therefore (unsurprisingly) one cannot ignore the nearby-bin mode coupling induced due to the mask.

\begin{figure}
    \centering
    \includegraphics[width=0.9\columnwidth]{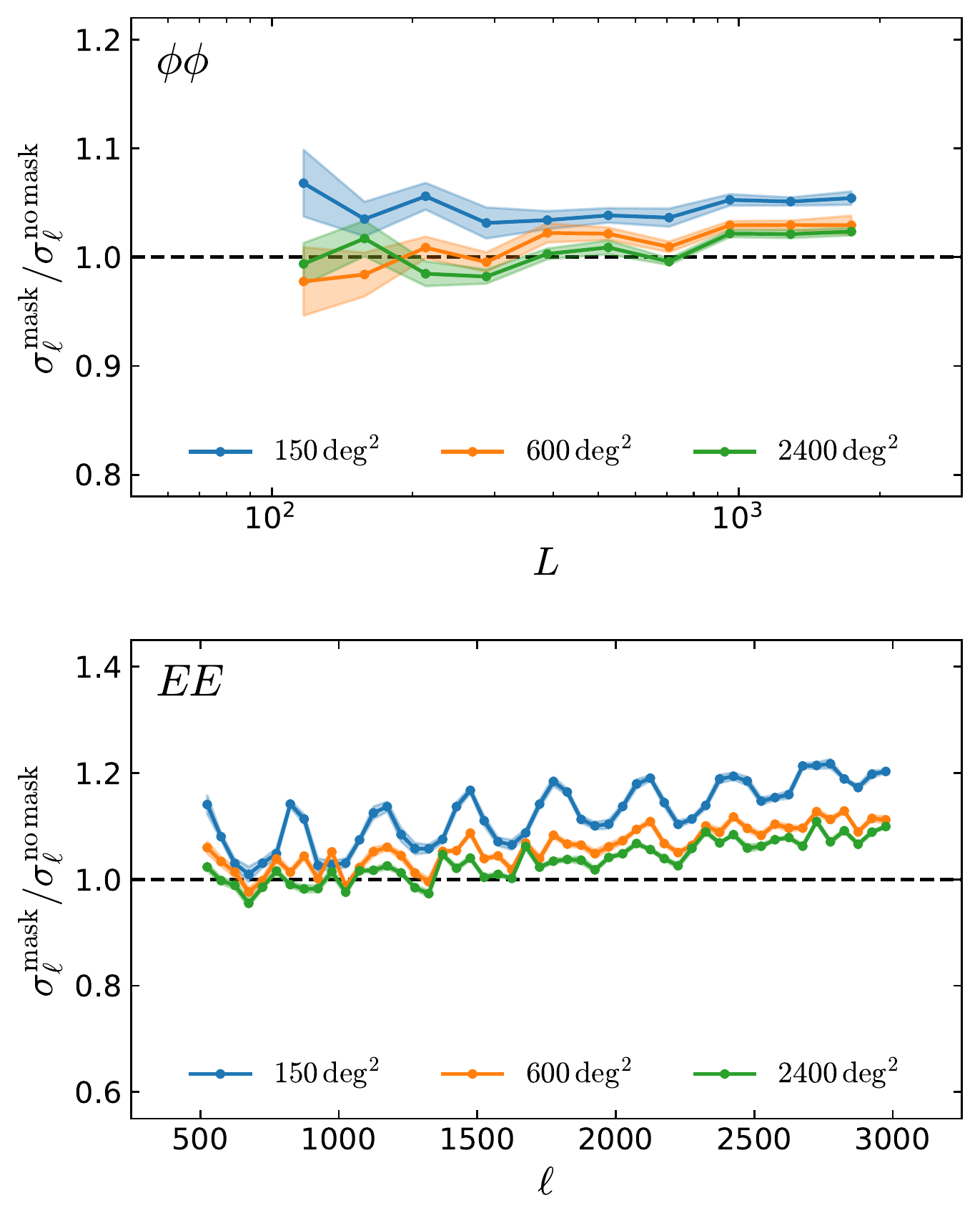}
    \caption{Demonstration of how masking leads to excess variance in inferences of lensing bandpowers and delensed $E$ mode bandpowers as compared to expectation from a simplistic $f_{\rm sky}$-scaled periodic sky forecast. Each panel shows the ratio of error bars computed from $\surd{\rm diag}(\Sigma^{\rm \MUSE})$ between a masked case and a periodic sky case with identical effective $f_{\rm sky}$. Error bands show the 1\,$\sigma$ standard error on these quantities estimated via bootstrap resampling, as the covariances are themselves computed via Monte Carlo. As the size of the field grows and the mask boundary is a smaller percentage of the total field, any differences reduce. By a 2400\,deg$^2$ (close to the smallest planned CMB-S4 field), the impact is ${\lesssim}\,10\%$, providing important confirmation of such forecasts even if masking was ignored.}
    \label{fig:mask_impact}
\end{figure}

We also wish to verify empirically that the joint estimate of $(\Aphib,\AEb)$ is unbiased and that its covariance, $\Sigma^{\rm MUSE}$, calculated based on Eqn.~\eqref{eq:JH}, accurately represents the empirical scatter of the estimate. This is demonstrated in Fig.~\ref{fig:bias_variance.pdf}. Colored lines show a suite of simulated joint estimates, and the solid and dashed lines show that their mean recovers the input theory model on average, and that their scatter matches the prediction based on $\Sigma^{\rm MUSE}$, respectively. This is the case whether masking is used or not (top and bottom panels). 

Thus far, all forecasting results, including all performed for CMB-S4, have only approximately accounted for the effects of pixel masking by assuming that constraints scale with $\sqrt{f_{\rm sky}}$, where $f_{\rm sky}$ is the fraction of the total sky which is observed. While this is expected to be reasonably accurate, particularly for small-scale power spectrum estimation, the story is more complicated for lensing reconstruction and delensing, where masking mixes some of the same lensing modes which in turn also impact delensing. Using \MUSE, we perform an important check of the accuracy of the $f_{\rm sky}$ scaling assumed in these forecasts. 

Specifically, we compare the \MUSE covariance for two cases, here in the \configSMALL configuration. In one case we apply a pixel mask and in the other we shrink the field so that it is the same effective $f_{\rm sky}$ as the masked case, but otherwise do not apply any masking. In Fig.~\ref{fig:mask_impact}, we show the ratio of $\surd {\rm diag}(\Sigma^{\rm MUSE})$ for the two cases. Additionally, we also scale up the size of the field to larger more realistic sizes (while keeping the size of the 1\degree\ border mask unchanged). We expect that as we increase the field size, any masking effects beyond $f_{\rm sky}$ should reduce, as the masking impacts a smaller and smaller fraction of the total modes which enter the estimate. This is indeed what we find, showing that while for the smallest ${\sim}\,150\,{\rm deg}^2$ field, an $f_{\rm sky}$-scaled forecast could be as large as 10\% too optimistic in the lensing reconstruction error bars and 20\% optimistic in the delensed $E$ mode error bars at high-$\ell$, by the time we reach realistic field sizes of a few thousand square degrees, the impact is almost nothing for $\phi$ and ${\lesssim}\,10\%$ for delensed $E$. This is an encouraging result which suggests masking will not pose any unexpected problems for lensing analyses of CMB-S4 data. In the future, it will be interesting to compare full-sky methods like those of \cite{green2016,hotinli2021} directly to \MUSE to quantify the impact of realistic data effects even further.

\subsection{Realistic example on SPT-3G data}
\label{sec:bigsky}

\begin{figure}
    \centering
    \includegraphics[width=\columnwidth]{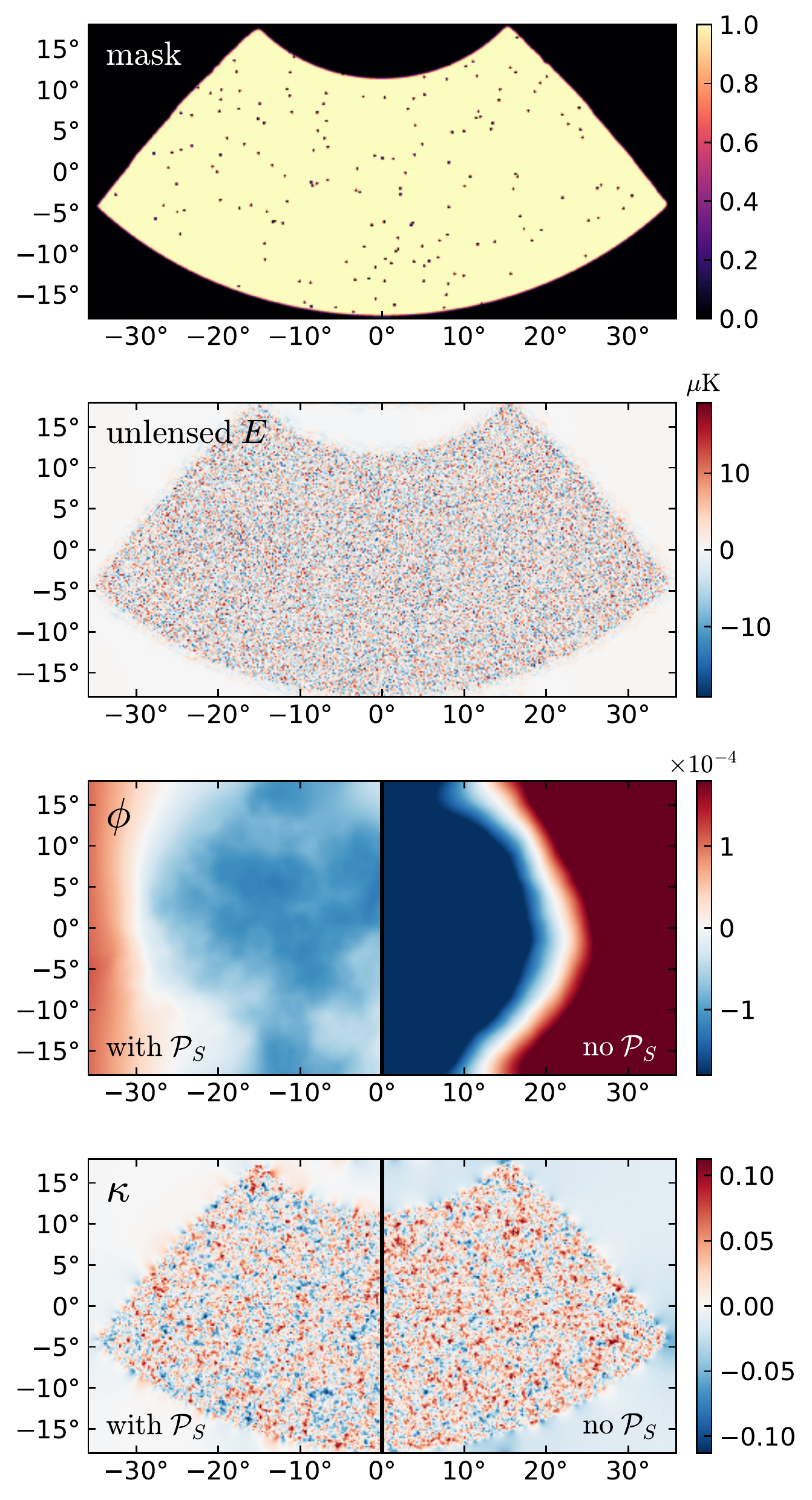}
    \caption{(Top panel) The pixel mask used in the analysis presented in Sec.~\ref{sec:bigsky}, which features 1500\,deg$^2$ of simulated SPT-3G data. (Next two panels) Typical joint MAP estimates of unlensed $E$ and $\phi$, which enter the \MUSE estimate as the point in the latent parameter space at which gradients with respect to CMB and lensing bandpowers are taken. (Bottom panel) The $\kappa\,{=}\,-\nabla^2\phi/2$ corresponding to the $\phi$ panel. In the bottom two panels, the left half of the image shows the result when imposing the super-sample prior (Eqn.~\ref{eq:supersample}), whereas the right half shows the result without any additional prior. The impact of this prior is to reduce the mean-field feature to levels small enough (though not necessary zero) as to render the \MUSE estimate unbiased and near-optimal.}
    \label{fig:MAP_maps}
\end{figure}
    
\begin{figure*}
    \centering
    \includegraphics[width=\textwidth]{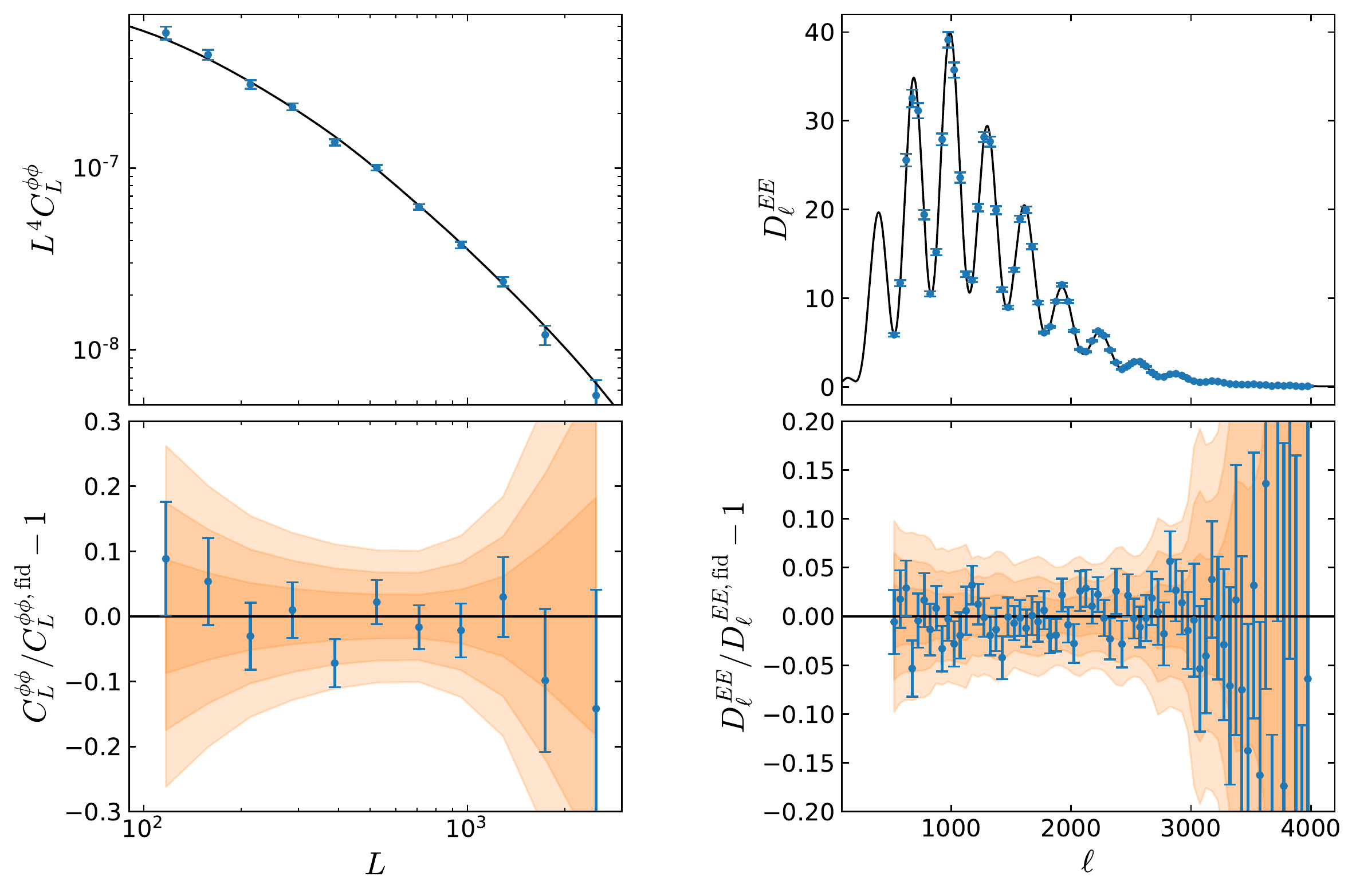}
    \caption{\MUSE estimates of the lensing potential and unlensed $E$-mode bandpowers for a 1500\,deg$^2$ simulated SPT-3G dataset. The black line shows the fiducial theory. The top panels show the absolute estimate, while the bottom panels show the fractional difference against the truth in units of percent. The orange shaded regions in the bottom panel are 1, 2, and 3 $\sigma$ error bands (they are the same as the size of the error bars, but easier to compare across different multipoles). This full analysis required around one hour of computing time, a drastic improvement over HMC sampling.}
    \label{fig:3g_spectra}
\end{figure*}

The previous section examines the properties of the \MUSE lensing estimate on a relatively small patch of sky where large suites of simulated analyses are computationally convenient. We now demonstrate the feasibility of \MUSE on a much larger region of sky, representative of the deep fields of ongoing and upcoming CMB surveys. Specifically, we will consider a 1500\,deg$^2$ region with the noise levels and masking expected for the upcoming SPT-3G survey \cite[][specification given in Table.~\ref{table:configurations}]{benson2014}. This is also similar in size to the region expected to be probed (to deeper noise levels) with the CMB-S4 survey for the main purpose of primordial gravitational wave detection \cite{abazajian2016}. We continue to work in the flat-sky approximation, which begins to break down near patches of sky of this size. We will comment later on a simple way to account for this within the \MUSE procedure. 

For visualization, Fig.~\ref{fig:MAP_maps} shows a typical SPT-3G pixel mask, as well as typical $\hat f_J$, $\hat \phi_J$, and $\hat \kappa_J\,{=}\,-\nabla^2 \hat\phi_J/2$ maps (only the $E$ component of $\hat f_J$ is shown since we do not estimate bandpowers of $B$, but the MAP reconstruction of $B$ is also non-zero and used in the algorithm). In the bottom two panels, the left half of the image shows the result when imposing the super sample prior defined in Eqn.~\eqref{eq:supersample}, whereas the right half show the result without this prior. The impact of the prior is to reduce the mean-field feature, which is visible as a large residual in $\phi$, and an additive offset in $\kappa$ (note the negative average $\kappa$ in the masked regions, and the positive average $\kappa$ in the unmasked region, barely visible as a slight preponderance for hot spots as opposed to cold spots there). The mean-field is a real feature of the joint MAP and arises due to any masking or filtering that breaks statistical isotropy or introduces non-Gaussianity (it is also present in the masked \configSMALL analyses in the previous subsections, and not unique to the SPT-3G mask). It poses problems for the \MUSE estimate as it leaks a large number of unconstrained modes into bandpowers of interest. Put another way, a $\kappa$ offset corresponds to an overall magnification which brings modes from outside of the mask into view, but which are unconstrained by data and hence lack the asymptotic guarantees needed by \MUSE. The super sample prior stops enough of these modes from entering and potentially leading to biases. We also note that the removal is not perfect---nor does it need to be---to render \MUSE effectively unbiased and near-minimum variance, as demonstrated in the previous subsection.

Computing estimates of $\hAphib$ and $\hAEb$ for the SPT-3G simulation proceeds identically as in the previous subsection. A typical MAP solution without a starting guess takes around two minutes to complete on a Tesla A100, and the entire \MUSE estimate is under an hour on a handful of GPUs. By contrast, a naive scaling of existing HMC methodology would predict over a week for a sufficiently converged chain. An estimate and error-bands for one particular simulation are shown in Fig.~\ref{fig:3g_spectra}. 

\section{Conclusion}
\label{sec:conclusion}

In this paper we have described the \MUSE algorithm, a generic method for hierarchical Bayesian inference. \MUSE is based on an approximation to the marginal score which is extremely fast to compute relative to exact methods. It is most applicable to problems where one needs to marginalize over a very high-dimensional latent space, but the final constraints of interest involve a small number of fairly well-constrained parameters. In the limit of a perfectly Gaussian latent space, \MUSE becomes exact and is equivalent to the MMLE. It also performs extremely well on funnel problems, which are challenging for other methods such as HMC. The only technical requirements of the \MUSE estimate are that one has access to gradients of the joint likelihood function and one can generate simulations from the forward model. This requirement is usually satisfied for a broad range of problems, including the common case of problems defined via probabilistic programming languages. In its current form, \MUSE is interpretable as either a frequentist estimator for parameters of interest which is asymptotically normal and unbiased, or as an approximate Bayesian procedure yielding Gaussianized marginal posterior inferences. 

Computing the \MUSE estimate is straight forward. To summarize, it involves:
\begin{enumerate}
    \item Finding the MAP estimate of the latent space variables, $z$, given a starting guess for $\theta$, and computing the gradient with respect to $\theta$ at this $z$. 
    \item Computing this same quantity on suite of data simulations from the forward model generated given the current $\theta$, and differencing the average of these from the data value. 
    \item Iteratively solving for where this difference is zero, yielding the \MUSE estimate. 
    \item Computing the $H$ and $J$ matrices according to Eqn.~\eqref{eq:JH}, then combining them to form the covariance, which involves only more MAP gradients of the same kind used in the estimate itself (and which can be sped up if second order AD is available).
\end{enumerate}

One can view the effectiveness of \MUSE as stemming from having replaced the difficult problem of high-dimensional integration with the much more tractable problem of high-dimensional optimization. In this sense, it is similar in spirit to both VI and EM, although differs from VI in that no surrogate distributions need to be chosen by the user. Of course, the ability to choose a well-tailored surrogate distribution is, in another sense, a strength of VI which makes it more generic. However, for VI to be tractable in high-dimensions, the surrogate distribution must often necessarily be a Gaussian, or in the case of mean-field VI, an uncorrelated Gaussian. Conversely, \MUSE implicitly deals with all correlations in the latent space, and does not necessarily correspond to a Gaussian latent space approximation. One can also view \MUSE as an approximate EM procedure, where the averaging over simulations in Eqn.~\eqref{eq:smuse} serves as the expectation step, and the root-finding over $\theta$ is analogous to the maximization step. \MUSE is also similar, but not equivalent, to the Laplace approximation, differing in that the Hessian of the latent space never needs to be computed. In general, \MUSE excels on high-dimensional problems because no dense high-dimensional operators are ever needed. In a more general sense, one could consider \MUSE a form of SBI, as it requires only the ability to generate forward simulations, along with the ability to compute gradients of the joint likelihood, which is often automatically satisfied if forward simulations are available. 

% We envision several extensions to \MUSE which may be possible to improve the accuracy of the approximation. Because \MUSE is exact for Gaussian latent spaces, a powerful combination may be to train a normalizing flow which Gaussianizes the latent space \cite{}, then use \MUSE to very quickly perform the latent marginalization, which will now be essentially exact. We are also exploring ways to use \MUSE to generate samples from from an approximate $\mathcal{P}(\theta\,|\,x)$ implicitly defined by \MUSE, the challenge there being that the non-conservative nature of $\smuse$ does not uniquely define such a posterior. Finally, it would be worth exploring whether it is advantageous to consider \MUSE as a summary statistic and apply SBI methods such as Approximate Bayesian Computation \cite{} or neural density estimation to the posterior $\mathcal{P}(\theta\,|\,\hatthetamuse)$, with \MUSE having served as a near-optimal massive reduction in the dimensionality of original joint likelihood.

We then applied \MUSE to the problem of estimating the unlensed CMB power spectra and gravitational lensing potential power spectra from realistic CMB data. The original works demonstrating that improvements over the standard QE procedure are possible \cite{hirata2003,hirata2003a} have motivated many followups attempting to render the method ready to apply to data. To consider the optimal lensing data analysis problem solved, we view it as necessary to demonstrate 1) joint estimation of bandpowers or parameters controlling both the lensing potential and the unlensed CMB spectra 2) proof that the estimate is optimal, for example by comparing against exact posterior distributions or against exact Fisher matrix calculations 3) demonstration that the procedure works and remains optimal even in the presence of masking and 4) is reasonably fast. Some methods have satisfied a subset of these requirements, but thus far \MUSE is the first to satisfy them all. Coupled with its conceptual simplicity, we therefore believe it is a very promising path forward for CMB lensing and delensing analysis.

Many CMB cosmologists intuitively view optimal lensing analysis as performed by an ``iterative quadratic estimate,'' based on popular discussion in \cite{smith2012}. Although there is no unique definition of the iterative QE, and the original discussion was only a heuristic forecasting procedure, it is true that many optimal lensing results take the form of iterating an estimate that is quadratic in the data. \MUSE can be afforded this interpretation too, since each gradient step used in obtaining the joint MAP that is part of the calculation is quadratic in the data, and this is iterated until convergence. The additional pieces of \MUSE lensing, mainly debiasing the score with simulations, performing the root-finding iterations, and the covariance prescription, can be considered as implicitly performing a cosmology-dependent power spectrum debiasing and noise quantification which is sometimes imagined as part of what a map-level iterative QE would entail. 

Our results here demonstrated that \MUSE is effectively unbiased and optimal for the noise levels and sky areas of all upcoming surveys, and in the presence of masking. A key but simple development which allowed \MUSE to work on CMB lensing was the addition to the super-sample prior discussed in Sec.~\ref{sec:lensingposterior}, which reduces the mean-field feature in the joint MAP, but is otherwise very non-informative in terms of the marginal posterior. We showed that at CMB-S4 noise levels, correlations between bandpowers of $\phi$ and bandpowers of unlensed $E$ can be ignored with minimal impact on resulting parameters even when masking is present. However, we caution that we did not consider higher noise levels nor correlations with lensed $E$ estimates which are known to be larger \cite{hotinli2021}. We also compared simple $f_{\rm sky}$-scaled forecasts given CMB-S4 noise levels with ones more exactly accounting for pixel masking. We found that $f_{\rm sky}$-scaled forecast are overly optimistic by as much as 20\% given very small patches of sky, but this reduces to ${\lesssim}\,10\%$ for larger realistic fields like the SPT-3G survey or even the smallest CMB-S4 fields. This is an assuring confirmation of forecasted next-generation constraints. 

Our code for this work assumed the flat-sky approximation, but the \MUSE algorithm is generic and directly generalizes to the curved sky as well. Although HMC sampling on the curved sky is at present slightly out of reach, \MUSE is much faster and runs more easily on the curved sky, and we expect to update the software accordingly in the near future. It is interesting to note, however, that there is an alternate and easier way to deal with sky curvature. The proof that \MUSE is asymptotically unbiased does not rely on $\smap$ being exact. Instead, the only requirement is that the simulations which are averaged over in $\smuse$ accurately describe the data distribution. Thus, as long as the forward simulations are generated on the curved sky, the much-costlier MAP solutions can still be computed on the flat-sky. This will not add any additional bias to the \MUSE estimate, only potentially excess variance, which will be captured in the $H$ and $J$ matrices. This can serve as a fast substitute for intermediate sky areas where the flat-sky approximation fails but any excess variance is still small. Additionally, the same argument allows us incorporate any of a number of other effects, such as foregrounds or other instrumental effects, in the simulations, but not in the MAP calculation, and still recover unbiased results. Future extensions can proceed by first incorporating these effects (possibly with free parameters which we will then infer) in the simulation model, and only accounting for them in the MAP when the excess variance is deemed too large. 

\begin{acknowledgements}
This material is based upon work supported by the National Science Foundation under Grant Numbers 1814370 and NSF 1839217, by NASA under Grant Number 80NSSC18K1274, and by the U.S. Department of Energy, Office of Science, Office of Advanced Scientific Computing Research under Contract No. DE-AC02-05CH11231 at Lawrence Berkeley National Laboratory. This research used resources of the National Energy Research Scientific Computing Center, which is supported by the Office of Science of the U.S. Department of Energy under Contract No. DE-AC02-05CH11231. We thank Emmanuel Schaan, Vanessa B\"ohm, and Selim Hotinli for useful discussions, and Ethan Anderes for the particular interpretation of the mean-field given in Sec.~\ref{sec:bigsky}.
\end{acknowledgements}

\onecolumngrid
\appendix

\section{\MUSE is exact for a Gaussian latent space}
\label{app:exactmpm}

The \MUSE gradient in Eqn.~\eqref{eq:smuse} is approximate in general, but
is exact if the Hessian is constant, which requires a Gaussian
likelihood with linear dependence of the data model on the latent
space, as well as a Gaussian prior. A typical case where this arises is when estimating parameters which control a signal covariance in the ``Wiener filter'' problem. To help build intuition, here we
demonstrate explicitly that \MUSE gives an exact answer to this problem, which also has an analytic solution. 

Consider the case where some data, $\x$, is the sum of signal and
noise, $s$ and $n$, both of which are Gaussian with covariances
$\op{S}(\theta)$ and $\op{N}$, respectively, the former depending on
parameters which we wish to infer, $\theta$. The signal $s$ plays the
role of latent variables $z$, i.e. these are the variables that need
to be marginalized over. In this case, the marginal is analytic and
the posterior distribution for $\theta$ is
\begin{align}
    2\log \mathcal{P}(\theta\,|\,\x) \propto -\x^\dagger \big( \op{S}(\theta)+\op{N} \big)^{-1}  \x
     - \log\det\big(\op{S}(\theta)+\op{N}\big)
\end{align}
The gradient of the RHS, which can be used to iteratively step to the MMLE, is
\begin{align}
    \label{eq:toyg}
    2g = \frac{d}{d\theta} \left[ -
    \x^\dagger \big( \op{S}(\theta)+\op{N} \big)^{-1}  \x  - \log\det\big(\op{S}(\theta)+\op{N}\big) \right] &= \\
    \x^\dagger \big( \op{S}(\theta)+\op{N} \big)^{-1} &\frac{d\op{S}}{d\theta} \big( \op{S}(\theta)+\op{N} \big)^{-1} \x - {\rm tr} \left[ \big( \op{S}(\theta)+\op{N} \big)^{-1} \frac{d\op{S}}{d\theta} \right].
\end{align}
Next, consider the \MUSE gradient from Eqn.~\eqref{eq:smuse} for the same problem. This uses the joint distribution of $\x$ and $s$,
\begin{align}
    \label{eq:toyjointposterior}
    2\log \mathcal{P}(s,\x\,|\,\theta) = -\frac{(d-s)^2}{\op{N}} - s^\dagger\op{S}(\theta)^{-1}s - \left[\log \det \op{S}(\theta)+\log \det \op{N}+(n_s+n_d)\log 2\pi\right].
\end{align}
The MAP estimate at fixed $\theta$, recognizable as the Wiener filter of the data, is
\begin{align}
    \label{eq:toyMAP}
    \hat s_{{\rm MAP}|\theta} = \underset{s}{\rm argmax} \; \log \mathcal{P}(s,\x\,|\,\theta) = \op{S}(\theta) \big(\op{S}(\theta)+\op{N}\big)^{-1} \x
\end{align}
The first term in Eqn.~\eqref{eq:smuse} is the gradient of Eqn.~\eqref{eq:toyjointposterior} evaluated at the MAP estimate from Eqn.~\eqref{eq:toyMAP}. This is
\begin{align}
    \frac{d}{d\theta} \left.\left[ -\frac{(\x-s)^2}{\op{N}} - \frac{s^2}{\op{S}(\theta)} - \log \det \op{S}(\theta)\right]\right|_{\hat s_{{\rm MAP}|\theta}} &= \left.\left[ s^\dagger \op{S}(\theta)^{-1} \frac{d\op{S}}{d\theta} \op{S}(\theta)^{-1} s - {\rm tr} \; \op{S}(\theta)^{-1} \frac{d\op{S}}{d\theta} \right]\right|_{\hat s_{{\rm MAP}|\theta}} \\
    \label{eq:toygmpmterm1}
    &= \x^\dagger \big( \op{S}(\theta)+\op{N} \big)^{-1} \frac{d\op{S}}{d\theta} \big( \op{S}(\theta)+\op{N} \big)^{-1} \x - {\rm tr} \; \op{S}(\theta)^{-1} \frac{d\op{S}}{d\theta}
\end{align}
The second term in Eqn.~\eqref{eq:smuse} subtracts the average of Eqn.~\eqref{eq:toygmpmterm1} over data, $\x$. Since the trace in Eqn.~\eqref{eq:toygmpmterm1} does not depend on data, this piece will cancel. The remaining piece is, 
\begin{align}
    \left \langle \x^\dagger \big( \op{S}(\theta)+\op{N} \big)^{-1} \frac{d\op{S}}{d\theta} \big( \op{S}(\theta)+\op{N} \big)^{-1} \x \right \rangle_{\x \sim \mathcal{N}(0, \op{S}(\theta)+\op{N})} = {\rm tr} \left[ \big( \op{S}(\theta)+\op{N} \big)^{-1} \frac{d\op{S}}{d\theta} \right]
\end{align}
where we have made use of the identity that ${\rm tr}\,\op{A}=\langle z^\dagger \op{A} z \rangle_{z\sim\mathcal{N}(0,\op{I})}$. This gives a full \MUSE gradient of
\begin{align}
    2\smuse = \x^\dagger \big( \op{S}(\theta)+\op{N} \big)^{-1} &\frac{d\op{S}}{d\theta} \big( \op{S}(\theta)+\op{N} \big)^{-1} \x - {\rm tr} \left[ \big( \op{S}(\theta)+\op{N} \big)^{-1} \frac{d\op{S}}{d\theta} \right]
\end{align}
We see this is identical to Eqn.~\eqref{eq:toyg}, confirming that in this Gaussian case, \MUSE gives the exact gradient and hence gives the exact MMLE upon iteratively stepping in the gradient direction. The key feature of \MUSE, made explicit in this example, is the use of Monte-Carlo to compute the gradient of the log-determinant which appears in the marginal posterior of $\theta$, phrased in a conceptually straightforward way which only requires computing MAP estimates and gradients of the joint posterior of $\theta$ and $s$.

It is also useful to show that marginal over the latent space is required and that MLE/MAP fails in these examples where the dimensionality of latent space equals that of the data. One may for example think that evaluating MAP+MLE of $s$ and $\theta$ simultaneously would be sufficient. One can see from equation \ref{eq:toyjointposterior} that maximizing all at the same time leads to solution $s^2=\op{S}=0$, such that $s^2 \propto \op{S}$, which is clearly the wrong solution. This is because MLE/MAP estimators are only asymptotically unbiased, which is not satisfied if $n_d=n_s$. Once we marginalize out $s$ we are left with $n_{\theta}$ parameters where $n_{\theta}\ll n_d$, and we can use MLE in the asymptotic limit.

\section{Simplification of $J$ and $H$ computation}
\label{app:JH}

In this appendix we show how to simplify the $J$ and $H$ matrices defined in Eqs.~\eqref{eq:defJ} and \eqref{eq:defH} to their final form in Eqn.~\eqref{eq:JH}. The $J$ matrix is defined in Eqn.~\eqref{eq:defJ} as
\begin{align}
    J_{ij} &= \Big\langle \smuse[i](\theta^\ast,\{\x_n\}) \, \smuse[j](\theta^\ast,\{\x_n\}) \Big\rangle_{\x_n\overset{\rm iid}{\sim}\,\mathcal{P}(\x\,|\,\theta^\ast)},
\end{align}
where 
\begin{align}
    \smuse[i](\theta^\ast,\{\x_n\}) = \frac{1}{N}\sum_{n=1}^{N} s_i^{\rm MAP}(\theta^\ast,\x_n) - \Big \langle s_i^{\rm MAP}(\theta^\ast,\x^\prime) \Big \rangle_{\x^\prime \sim \mathcal{P}(\x^\prime\,|\,\theta^\ast)}.
\end{align}
Substituting this in and dropping the arguments to the summation, to $\smuse$, and to the expectation value for clarity yields,
\begin{align}
    \bigg\langle \bigg( \tfrac{1}{N}\textstyle\sum s_i^{\rm MAP} - \Big \langle s_i^{\rm MAP} \Big \rangle \bigg) \bigg( \tfrac{1}{N}\textstyle\sum g_j^{\rm MAP} - \Big \langle g_j^{\rm MAP} \Big \rangle \bigg) \bigg\rangle 
    \;\; \rightarrow \;\; \Big\langle \smap[i] \, \smap[j] \Big\rangle -\Big\langle \smap[i] \Big\rangle \Big \langle \smap[j] \Big\rangle
\end{align}
where in the second step we have taken the limit $N\rightarrow\infty$. This yields the result in Eqn.~\eqref{eq:JH} (which is written there with the arguments included).

The $H$ matrix is defined in Eqn.~\eqref{eq:defH} as:
\begin{align}
    H_{ij} &= \Big \langle \frac{d\smuse[i]}{d\theta_j} (\theta^\ast,\{\x_n\}) \Big \rangle_{\x_n\overset{\rm iid}{\sim}\,\mathcal{P}(\x\,|\,\theta^\ast)}.
\end{align}
Substituting and simplifying, we find
\begin{align}
    & \Big\langle \frac{1}{N}\sum_{n=1}^{N} \frac{ds_i^{\rm MAP}}{d\theta_j}(\theta^\ast,\x_n) \Big\rangle_{\x_n\overset{\rm iid}{\sim}\,\mathcal{P}(\x\,|\,\theta^\ast)}  - \left. \frac{d}{d\theta_j} \Big \langle s_i^{\rm MAP}(\theta,\x^\prime) \Big \rangle_{\x^\prime \sim \mathcal{P}(\x^\prime\,|\,\theta)} \right|_{\theta=\theta^\ast} \\
    &= \Big\langle \frac{ds_i^{\rm MAP}}{d\theta_j}(\theta^\ast,\x) \Big\rangle_{\x\sim\mathcal{P}(\x\,|\,\theta^\ast)}  - \left.  \frac{d}{d\theta_j} \left[ \Big \langle s_i^{\rm MAP}(\theta,\x^\prime) \Big \rangle_{\x^\prime \sim \mathcal{P}(\x^\prime\,|\,\theta)} \right] \right|_{\theta=\theta^\ast} \label{eq:Hsimplify} \\
    &= - \left. \frac{d}{d\theta_j} \left[ \Big\langle \smap[i](\theta^\ast,\x) \Big\rangle_{\x\sim\mathcal{P}(\x\,|\,\theta)} \right] \right|_{\theta=\theta^\ast}
\end{align}
%`'
where the last line follows from noting that there are two chain rule terms arising from the term in brackets in Eqn.~\eqref{eq:Hsimplify}:  one where the derivative acts on the $\theta$ inside the expectation value, and another when it acts on the $\theta$ controlling the distribution over which the expectation value is taken. The first of these chain rules terms is canceled by the first the term in Eqn.~\eqref{eq:Hsimplify}, leaving only the second, which yields the result in Eqn.~\eqref{eq:JH}.

\bibliography{marius}

\end{document}